\documentclass[aps,superscriptaddress,showpacs,twocolumn,prx,nobibnotes]{revtex4-1}
\usepackage[colorlinks=true,citecolor=blue]{hyperref}
\usepackage{graphicx}
\usepackage{epstopdf,epsfig}
\usepackage{amsmath}
\usepackage{amsfonts}
\usepackage{epstopdf,epsfig}
\usepackage{times}
\usepackage{amssymb}

\usepackage{color}
\usepackage[normalem]{ulem}
\usepackage{braket}
\usepackage{xcolor}

\newcommand{\beq}{\begin{equation}}
\newcommand{\eeq}{\end{equation}}
\newcommand{\bea}{\begin{eqnarray}}
\newcommand{\eea}{\end{eqnarray}}

\renewcommand{\b}[1]{\mathbf{ #1}}

\newcommand{\Tr}{\textrm{Tr}}

\graphicspath{{figure/}}

\usepackage{mathtools}

\begin{document}

\title{Long-range topological insulators and weakened bulk-boundary correspondence}

\author{L. Lepori\email[correspondence at: ]{llepori81@gmail.com}}
\email[correspondence at: ]{llepori81@gmail.com}
\affiliation{Dipartimento di Scienze Fisiche e Chimiche, Universit\`a dell'Aquila, via Vetoio,
I-67010 Coppito-L'Aquila, Italy}
\affiliation{INFN, Laboratori Nazionali del Gran Sasso, Via G. Acitelli, 22, I-67100 Assergi (AQ), Italy.}

\author{L. Dell'Anna}
\affiliation{Dipartimento di Fisica e Astronomia, Universit\`a di Padova, via Marzolo 8, I-35131 Padova, Italy.}

\begin{abstract}

{
\color{black}
We  investigate the appearance of new types of insulators  and superconductors in long-range fermionic quantum systems. These phases are not included in the famous "ten-fold way classification'', valid in the short-range limit. This conclusion is obtained analyzing at first specific one-dimensional models, in particular their phase diagrams and entanglement properties. The long-range phases are signaled, for instance, by the violation of the area-law for the Von Neumann entropy and by a corresponding peculiar entanglement spectrum. 
Later on,  the origin of the deviations from the ten-fold way classification is investigated from a more general point of view and in any dimension, showing that it is 
related with the presence of divergences occurring in the spectrum, due to the long-range couplings. 
A satisfying characterization for the long-range phases can be achieved, at least for one-dimensional quantum systems, 
as well as the definition of a nontrivial topology for them, {\color{black} resulting in the presence of massive edge states,} provided a careful evaluation of the long-range contributions.
Our results allows to infer, at least for one-dimensional models, the weakening of the bulk-boundary correspondence, due to the important 
correlations between bulk and edges,  and consequently to clarify the nature of the massive edge states.  The emergence of this peculiar edge structure is signaled again by the bulk entanglement spectrum. 
The stability of the long-range phases against local disorder
is also discussed, showing notably that this ingredient can even strengthen the effect of the long-range couplings. 
 Finally, we analyze the
 entanglement content of the paradigmatic long-range Ising chain, inferring again important deviations from the short-range regime, 
as well as the limitations of bulk-boundary (tensor-network based) approaches to classify long-range spin models.
}
\end{abstract}
\maketitle

\section{Introduction} 
\label{intro}

The study of topological phases of matter
experienced a growing interest in the last decades.  
In the absence of interaction, a central result is the complete
classification of the topologically inequivalent (families of) phases for fermionic  systems, the famous
{\color{black} "ten-fold way classification" (TWC)} \cite{zirnbauer1996,zirnbauer1997,ludwig2008,kitaev2009,ludwig2009,ludwig2010}.
The systems included in this scheme host a "symmetry-protected topological order'', 
indeed their nontrivial topology is constrained and protected by some discrete symmetries,
oppositely to genuine topological order.
This theoretical achievement have been confirmed and corroborated   
 by the experimental characterization of solid-state 
 compounds with topological properties 
\cite{hasankane,zhang11,book_intro,chiu15}. \\
In spite of an energy gap obstructing in general charge or spin bulk conductivity, 
the main macroscopic property
exhibited by a nontrivial topological insulators {\color{black} and superconductors}
is the presence of edge conductivity, due to massless modes localized therein and well distinct
from bulk excitations. Moreover, phases with different topology are separated each others by continuous transitions, where the bulk mass gap vanishes.
Concerning the entanglement properties, 
the matter included in the ten-fold way classification displays short-range entanglement and correlations, the opposite situation holding again for genuine topological order \cite{wen2014}.

All the mentioned results are specific for quantum systems described by Hamiltonians
with short-range terms only. However, in the last years  
also the study of long-range classical and quantum systems \cite{libro}, both at and out 
of the equilibrium, gained a renewed attention.\\
Independent theoretical studies have shown that long-range quantum systems 
can exhibit various peculiar features, mostly stemming from the 
breakdown of lattice locality 
\cite{hast,hauke2013,eisert2014,metivier2014,noinf,damanik2014,nbound,
storch2015,carleo,kastner2015ent,Maghrebi2015,kuwahara2015,maghrebi2015-2,wouters2015,daley2016}. 
This set includes static correlation functions with hybrid (exponential 
and algebraic) decay \cite{deng2005,koffel2012,nostro,paperdouble}, 
anomalous growth for the entanglement after quenches \cite{growth},
new constraints on thermalization \cite{santos2015} and on conductivity in NS/NSN junctions \cite{giuliano2017}.\\
Even more interestingly, very recent works \cite{nostro,ares,paperdouble,delgado2015,
paper1,gori,gong2015,maghrebi2015-2,gong2015-2,lepori2016,alecce2017,neupert2016}
have suggested that long-range systems can host new phases at 
sufficiently small values of the decay exponents  $\alpha$ for the Hamiltonian terms.
These phases often manifest interesting features not owned by the short-range ones, including
continuous quantum phase transitions without mass gap closure, 
violation of the area-law for the Von Neumann entropy and of the Mermin-Wagner theorem,  emergence of 
massive edge states. \\
The occurrence of these 
properties, {\color{black} some of them also checked in experiments of trapped ions \cite{exp1,exp2},}  opened various issues and problems.
In \cite{nostro,paper1,lepori2016} it has been inferred that, for not interacting long-range lattice models, most of the described peculiarities  can be related with the action of some states in the bulk spectrum, called "singular states'', encoding some divergences related with the algebraic decay of the long-range couplings.\\
In spite of these important clues,
the understanding of the physical origin of the mentioned purely long-range phases, as well as
of their bulk and edge features, is still an open problem. Closely related, it appears a central issue to classify these phases,
understanding how the ten-fold way classification evolves in the presence of long-range Hamiltonian terms, when also correlation functions have been found not exponentially decaying any longer.

In the present paper we start to investigate this problem. Using first specific one-dimensional free fermionic examples and later on performing a more general formal discussion (not limited to one-dimensional cases), we show that long-range insulating {\color{black} or superconducting} phases can emerge, in some cases hosting massive edge states, when the bulk spectrum manifests a \emph{particular} sub-set of the mentioned singularities.  {\color{black} The appearance of the latter singularities parallels the area-law violation for the Von Neumann entropy,  still in the presence of a nonvanishing bulk mass gap, and a peculiar distribution for entanglement spectrum.}\\
{\color{black} We stress that, although our discussion exploits mainly superconducting models as specific examples, our results are not limited to them,
but concern also the strictly meant (topological) insulators. Indeed that possible appearance of the mentioned singularities in the spectrum does not depend directly on the superconducting or insulating nature of the bulk.\\
Due to the same singularities}, the definition of topology must be reconsidered ab initio, 
requiring a proper generalization of the approaches valid in the short-range limit.\\
Finally,  we infer, at least for one-dimensional systems, that the so-called bulk-boundary correspondence, typical of the short-range topological insulators {\color{black} and superconductors}, gets weakened in the long-range topological phases, as well as the definition itself of localized edge state valid in the short-range limit, due to the strong long-range correlations between the edges and with the bulk dynamics. {\color{black} Indeed a nontrivial LR topology still reflects in the presence of states localized on the edges, but these states have a nonzero mass and consequently a dynamics which is not separable from the one of the bulk (in the sense that no modes localized on a single edge can be defined from the bulk states), as happens instead in the short-range limit.}

Notably, some of the ideas and results achieved for one-dimensional long-range quadratic systems can hold, under specific restrictions, 
for higher-dimensional ones, as well as for interacting and/or spin long-range models.

The paper is organized as follows.  In Section \ref{kitaev} we recall at first two specific examples of one-dimensional
 fermionic long-range quantum systems, discussing their phase diagrams and ground state properties, with more details for algebraic long-range decay with exponent $\alpha<1$. {\color{black} Afterwards, starting from the analysis of previous results, in Section \ref{secbulk}
  we infer} that some gapped phases hosted by these systems do not insert in the classification for the short-range topological insulators and superconductors, but display a purely long-range nature. This thesis is reinforced in Section \ref{ES} by the analysis of the entanglement spectrum for the ground states after a spatial bipartition. This analysis is one of the main results of the present manuscript, as well as the discussion of its consequences, performed in Section \ref{failES}. In Section \ref{origin} we {\color{black} investigate at a more formal level the generic  inapplicability of the TWC when long-range Hamiltonian terms are added, 
 reconsidering the classification of the maps from the Brillouin zone induced by the Hamiltonian and nonlinear $\sigma$-model approaches to the TWC.} Notably, this discussion is again not limited to one-dimensional systems. In Section \ref{class} we deal, {\color{black} in part for the first time,} with the classification, by Berry phase and winding numbers, of the long-range phases encountered in the previous Sections,  as well as with the limitations and open problems concerning these approaches. At then end, we address the generalization of these methods to long-range free fermionic models with different symmetries and dimensionality. In Section \ref{failES}  we analyze at first the behaviour of the correlation length in long-range systems. Later on, starting from the latter discussion and from the results on the entanglement spectrum,  we infer the weakening of the bulk-boundary correspondence in the long-range topological phases, clarifying the nature of their massive edge states. {\color{black} This is another central result of the present paper.} In Section \ref{dissec} we discuss the stability of the long-range phases 
against  local disorder, expected to smear the effects of the long-range Hamiltonian terms.  In Section \ref{LRIsec} we probe the possible extension of some results obtained so far to other long-range models, spin-based and/or interacting. For this task, we {\color{black} analyze the entanglement content of the  paradigmatic long-range Ising model, finding again peculiarities in the entanglement spectrum} at small enough $\alpha$. Conclusions are finally presented in Section \ref{conclusions}. Further details, mentioned in the main text but not immediately required to understand it, are given in the Appendices \ref{kitaevapp}-\ref{appedge}.

\section{Discussion of previous results}
\label{kitaev}
In this Section we {\color{black} recall at first some basic features of the} two long-range (LR) generalization of the short-range (SR) Kitaev Hamiltonian \cite{kitaev}.
Further material is given in Appendix \ref{kitaevapp}.
{\color{black} Later on, analyzing the previous results about these chains, we infer that their phases at $\alpha<1$  cannot be included in the TWC for the SR topological insulators and superconductors \cite{zirnbauer1996,zirnbauer1997,ludwig2008,kitaev2009,ludwig2009,ludwig2010}.}

\subsection{The models}

In \cite{nostro,paper1,lepori2016} two quadratic quantum models involving  spinless fermions
on a one-dimensional lattice have been studies extensively. The first one is 
characterized by a LR pairing:
\begin{equation}
\begin{split}
H_{\mathrm{lat}} & = - w \sum_{j=1}^{L} \left(a^\dagger_j a_{j+1} + \mathrm{h.c.}\right)  - \mu \sum_{j=1}^L \left(n_j - \frac{1}{2}\right) +
\\ &+\frac{\Delta}{2} \sum_{j=1}^L \,\sum_{\ell=1}^{L-1} d_\ell^{-\alpha} \left( a_j a_{j+\ell} + a^\dagger_{j+\ell} a^\dagger_{j}\right) \, .
\label{Ham}
\end{split}
\end{equation}
For a closed chain, we define in Eq.~\eqref{Ham} $d_\ell = \ell$ ($d_\ell = L-\ell$) if $\ell < L/2$ ($\ell > L/2$) and 
we choose anti-periodic boundary conditions \cite{nostro}.

The spectrum $\lambda_{\alpha}(k)$ of the Hamiltonian in Eq.~\eqref{Ham} displays a critical line at $\mu = 1$ 
for every $\alpha$ and a critical semi-line $\mu = -1$ for $\alpha >1$.  Notably  the energy of the quasiparticles
diverges  in $k = \pi$ if $\alpha \leq 1$, while it displays, at every finite $\alpha$ and at the same momentum,  divergences  in some $k$-derivatives for $\lambda (k)$ (\cite{nostro,paper1}). For these reasons the states close to $k = \pi$ are called "singular states" (and their dynamics as "singular dynamics") \cite{lepori2016}; as mentioned in the Introduction they have shown responsible of the deviations from the SR behaviours, concerning for instance {\color{black} the phase content,} the decay of the static correlation functions, the breakdown of conformal symmetry at criticality and the underlying violation of the lattice locality. {\color{black} The stability of these features against finite-size effects, smearing the divergences of the singular states, is discussed in the Appendix \ref{finite}.} \\
Importantly,  at least for the closed chain, the ground state energy is still extensive also at $\alpha<1$, in spite of the singular states, so that no Kac rescaling is required \cite{libro,nostro}.

   For future purposes, it is convenient to report the tight-binding matrix Hamiltonian  corresponding to Eq. (\ref{Ham}):
\begin{equation}
H (k)= 
\begin{pmatrix}
-(w \cos k - \frac{\mu}{2})		&		{\color{black} -i} \frac{\Delta}{2} f_{\alpha} (k+\pi) \\
{\color{black} i} \frac{\Delta}{2} f_{\alpha} (k+\pi) 			& 		(w \cos k - \frac{\mu}{2})
\end{pmatrix}
\label{tbHam}
\end{equation}      
in the (momentum diagonal) space $\big(a_{k} ,  a^\dagger_{-k} \big)$. The function 
$f_{\alpha} (k) \equiv \sum_{\ell=1}^{L-1} \sin(k \ell)/d_\ell^\alpha$ 
{\color{black} is singular at $k=0$ in the thermodynamic limit, $L\rightarrow \infty$}, 
encoding the mentioned singularities from the LR character of the Hamiltonian.

The Hamiltonian in Eqs. \eqref{Ham} and \eqref{tbHam} shares the same symmetries of the SR Kitaev chain, that means, beyond the unitary $Z_2$ parity of the total fermionic number, the anti-unitary charge conjugation and the  time reversal symmetries. This content in symmetries and the properties of the operators realizing them formally locates the model in  Eq. (\ref{Ham}) in the class BDI of the TWC \cite{zirnbauer1996,zirnbauer1997,ludwig2008,kitaev2009,ludwig2009,ludwig2010}.\\

Some generalizations of the Hamiltonian in Eq. (\ref{Ham}), involving as well a LR hopping, can be also considered:
\begin{equation}
\begin{split}
H_{\mathrm{lat}} & = - w \sum_{j=1}^{L} \,\sum_{\ell=1}^{L-j} d_\ell^{-\beta} \left(a^\dagger_j a_{j+\ell} + \mathrm{h.c.}\right)  - \mu \sum_{j=1}^L \left(n_j - \frac{1}{2}\right) 
\\ & + \frac{\Delta}{2} \sum_{j=1}^L \,\sum_{\ell=1}^{L-j} d_\ell^{-\alpha} \left( a_j a_{j+\ell} + a^\dagger_{j+\ell} a^\dagger_{j}\right) \, .
\label{Ham2}
\end{split}
\end{equation}
These models have been studied in \cite{paperdouble,alecce2017}.
The structure and the expression for the energy of the ground-states is very similar to the ones for the Hamiltonian in Eq. (\ref{Ham}), with the difference in Eq. (\ref{tbHam}) {\color{black} that $\cos k \to -g_{\alpha} (k+\pi)$ (for $\beta = \alpha$)
\beq
g_{\alpha} (k) \equiv \sum_{\ell} \cos(k \ell)/d_\ell^\alpha \, .
\label{defg}
\eeq}
Still divergences in the quasiparticle spectrum occur at $k = \pi$ if $\alpha <1$ and also high-order ones at every finite $\alpha$. However these divergences display a central difference compared with the ones from $f_{\alpha}(k)$ in 
Eq. \eqref{tbHam}: indeed $g_{\alpha} (-k) = g_{\alpha} (k)$, while $f_{\alpha} (-k) = - f_{\alpha} (k)$, affecting differently the singular dynamics, 
as we will see in detail in Section \ref{origin}. 
Again, the Hamiltonian in Eq. (\ref{Ham2}) shares the same symmetries of the SR Kitaev chain, then it belongs to the BDI class of the TWC.

\subsection{The phase diagrams}
\label{phased}

Concerning the phase diagram of the Hamiltonian in Eq. \eqref{Ham}, in \cite{nostro} it has been found that above the line $\alpha = 1$ two phases take place (at $|\mu| <1$ and $|\mu| >1$ respectively), continuously connected with the ordered and disordered phase  of the SR Kitaev chain. {\color{black} There the area-law Von Neumann entropy after a bipartition is fulfilled, as common in SR {\color{black} gapped} systems \cite{plenio2010} (although exceptions are known in peculiar ad hoc constructed models, see e.g. \cite{shor2016}). This means in formulae that
\beq
S(l \to \infty) \to a \, ,
\eeq
where $l$ characterizes the two parts of the chain with length $l$ and $L-l$ (it also holds $L \to \infty$) and $a$ is a constant.} \\
At variance, below the line $\alpha = 1$ two phases appear (at $\mu <1$ and $\mu >1$), 
signaled by a  deviation from the area-law for the Von Neumann entropy.   {\color{black} In particular this deviation turns out to be ruled by a logarithmic scaling law, as for SR quantum systems at criticality \cite{wilczek,calabrese}. In this way, similarly to what done in \cite{koffel2012,nostro,paperdouble}, the area law violation can be modeled as follows
\beq
S(l) = \frac{c_{\mathrm{eff}}}{6} \, \mathrm{ln} \, l \, .
\label{dev}
\eeq }
 In Eq. \eqref{dev} a value $c_{\mathrm{eff}} \neq 0$ signals the area-law violation.
The same violation has been demonstrated in \cite{paper1} by an effective theory close to the critical (semi-)lines $\mu = \pm 1$.

{\color{black} The described phase diagram is shown,  adapted from \cite{nostro}, in the Appendix \ref{kitaevapp}.}

A further striking feature of the zone below $\alpha = 1$ is that  at $\mu <1$  massive states localized on the edges appear \cite{nostro}, remnant of the Majorana (massless) edge modes present if $|\mu| <1$ and $\alpha >1$  (in the SR limit they are proper of the ordered phase for the SR Kitaev chain \cite{kitaev}, for a review see also \cite{fendley2012}).  Indeed in  \cite{paperdouble}  an hybridization mechanism of the Majorana modes, yielding the massive edge states, has been conjectured, similar to the one occurring  at finite sizes in the SR limit  \cite{kitaev}.  
{\color{black} The same mechanism has been finally proven in the limit $\alpha \to 0$ in \cite{neupert2016}.}

{\color{black} Interestingly, the phases at $\alpha< 1$ are not separated by any mass gap closure, nor by any first-order transition (see the next Section), from the ones at $\alpha>1$ \cite{nostro}.
 However at least a LR phase at $\mu <1$ and $\alpha<1$ is required by the finiteness of the mass gap in the same range: if this phase were not present, it could be possible to interpolate continuously between the ordered and disordered  phases at $\alpha>1$ and $\mu<1$ (see the phase diagram recalled in the Appendix \ref{kitaevapp}).

Some discontinuities, suggesting phase transitions, around $\alpha =1$ have been observed, e.g., in the Von Neumann entropy at half chain and in the related  mutual information, in the Berry phase \cite{delgado2015}, in the fidelity susceptibility and in the finite-size scaling behaviour of the multipartite entanglement \cite{pezze2017}.

\subsection{
Deviations from TWC: clues from previous results
\label{secbulk}
}

The phases at $\alpha<1$ for  the fermionic Hamiltonian in Eq. \eqref{Ham}  cannot be included in the TWC for the SR topological insulators and superconductors \cite{zirnbauer1996,zirnbauer1997,ludwig2008,kitaev2009,ludwig2009,ludwig2010}. In favour of this thesis,  we identify some evidences, elaborating some results from previous works (mainly \cite{paper1,paperdouble,delgado2015}):\\

\emph{i)}  the appearance of massive edge states \cite{paperdouble,delgado2015} in itself already signals a breakdown of the TWC, where only massless edge modes are  expected.
 The origin of these modes will be clarified in Section \ref{failES}.\\

{\color{black}
\emph{ii)} the TWC does not consider continuous phase transitions (especially between different topologies) without mass gap closure, as in the present case approaching the line $\alpha = 1$ (see \cite{nostro} and the previous Section).

The possibility of a first-order phase transition seems ruled out in our cases by the absence of divergences in the first derivative in $\alpha$ of the extensive ground-state energy $e_0(\alpha, L)$. Similarly, a crossover is excluded 
by a recent investigation of the fidelity susceptibility \cite{pezze2017}, as well as 
by basic considerations about the ground-state structure of open chains: if $\alpha>1$ the vanishing mass of the Majorana edge modes at $|\mu|<1$ implies the existence of two degenerate ground-states with different $Z_2$ fermionic parity, while the nonzero edge mass at $\alpha<1$ reflects in a unique ground state with even parity.
The described ground state structure deeply affects also the entanglement (spectrum) content, as will be discussed in Section \ref{ES}.\\
The absence of a mass gap closure can be justified heuristically by the large algebraically decaying  tails of the static correlation functions at small values of $\alpha$, also in the presence of a nonvanishing mass gap \cite{nostro,paperdouble,paper1,notesingul}.\\
}

\emph{iii)} the TWC involves SR (exponentially decaying) correlations and the fulfillment of the area-law for the Von Neumann entropy between disconnected subsystems. However, for the Hamiltonian in Eq. \eqref{Ham}  below $\alpha = 1$, where the quasiparticle energy diverges, this law is violated,  even if only logarithmically \cite{nostro}.\\

\emph{iv)}  the winding numbers $w$, characterizing the phases of the Hamiltonian in Eq. \eqref{Ham} at every $\alpha$ at least above 1, can be calculated directly following \cite{tewari2012}:
\beq
w = \frac{1}{2 \, \pi} \, \int_{BZ} \, \mathrm{d} \theta_k  \quad , \quad \theta_k = \mathrm{arcsin} \,  \Bigg[\frac{f_{\alpha} (k)}{\lambda_{\alpha} (k)} \Bigg] \, .
\eeq
$f_{\alpha} (k)$ and $\lambda_{\alpha} (k)$ defined in Section \ref{kitaev}.\\
The results are correctly found $w=0$ and $w = 1$ at $\alpha > 1$. On the contrary, at $\mu <1$ a fake semi-integer winding number $w =\frac{1}{2}$ appears below $\alpha = 1$, while $w =-\frac{1}{2}$ is obtained at $\mu >1$ \cite{delgado2015}. In the latter phase no edge states if found, and both the phases  are characterized by a unique ground-state. These results signal a clear inconsistence in the definition of topology by the winding numbers valid in the SR limit, since these numbers, when properly defined, can assume only integer values \cite{volovik,nakahara}. 
However, the mere emergence of this inconsistence can be interpreted as a signal of TWC deviation, related with the other LR features described above, as we will detail in Sections \ref{origin} and \ref{class}.\\

Finally, a bit more subtle but very relevant argument  is given in \cite{notetwophases}.\\

A similar analysis can be performed for the Hamiltonians in Eq. (\ref{Ham2}), having the same symmetries of Eq. (\ref{Ham}). This analysis leads to qualitatively equal conclusions as the ones just above. Indeed in \cite{paperdouble}
for the case $\beta  = \alpha$ at $\alpha \lesssim 1$ again an extended region has been found in the phase diagram where massive edge states appear.
 Again this phase has a single ground-state, paralleling  the nonvanishing mass of the edge states; however the same phase is not continuously connected with the disordered phase of the SR Kitaev chain.    \\

 Summing up, the arguments given above yield a quite compelling evidence that the phases at $\alpha <1$ on the Hamiltonians in Eqs. \eqref{Ham} and \eqref{Ham2} escape the TWC for SR fermionic systems.\\
  
\section{Deviations from TWC: further evidences from entanglement spectrum}
\label{ES}

{\color{black} The violation of the area-law for the Von Neumann entropy {\color{black} in gapped regimes at $\alpha <1$, suggesting the appearance of new purely LR phases,  induces a deeper study of the entanglement content in the same regimes.
For this reason, in the present  Section} we analyze, {\color{black} for the first time to our knowledge,} the entanglement spectrum (ES) for the 
non-critical LR paired Kitaev chain, Eq. (\ref{Ham}). {\color{black} This study, which is one of the main results of the present paper,} will help us
to  determine in deeper detail the structure of the phases 
at $\alpha <1$, corroborating their purely LR nature and linking together their peculiar properties.

The ES is defined in general as the set of (Schmidt)-eigenvalues of the reduced density matrix $\rho_B$ of a part $B$ of the considered quantum system
after a bipartition (see e.g. \cite{peschel2011}). It is known that ES is encoding even more information 
than the Von Neumann entropy \cite{li2008,fik2010,dechiara2012,lepori2013} 
and it can be calculated following the techniques 
described in \cite{peschel2011}.  \\
We assume in particular an open chain with  total length $L$  and bipartite it in such a way to isolate a 
segment of it, say between $L/4$ and $3 L / 4$.\\
We find that, below $\alpha =1$ and in the gapped regime $\mu \neq 1$, in correspondence with 
the violation of area-law for the Von Neumann entropy, the ES
resembles the typical one of a SR model at a critical point, 
assuming indeed a nearly continuous distribution {\color{black} (see \cite{calabrese2008,dechiara2012} and references therein)}. {\color{black} In the light of this behaviour, it deserves future effort to probe if this distribution is reproduced by the law found for critical one-dimensional SR systems reported in \cite{calabrese2008}. There the relevant parameter appearing in the distribution law is the conformal charge $c$, while in our case the same role should be played by the effective parameter $c_{\mathrm{eff}}$ governing the area law violation for the Von Neumann entropy, according to Eq. \eqref{dev}.}   \\
More importantly, the degeneracies of the Schmidt eigenvalues are found {\color{black} to be different from the ones generally expected  for gapped SR systems with same symmetries, as we will see in what follows}, signaling the appearance of pure LR phases.
 
The explicit results are shown in Fig. \ref{fig:ES}. 
There it can be seen that for $\alpha \gtrsim 1$ the Schmidt eigenvalues 
$\omega_n$ ($n$ labelling them starting from the highest one) composing the ES
are arranged in well-separated multiplets.
In particular when $|\mu| <1$ (left panel, showing the case $\mu = -0.5$ and $\alpha=3$), the dimension of the multiplets is even, as implied by 
the presence of two degenerate vacua (in the thermodynamic limit) $|\mathrm{GS}\rangle$ and $|\mathrm{GS}\rangle_0$ with different $Z_2$ fermionic parity, as recalled in the Appendix \ref{kitaevapp}
(see also \cite{fik2010,turner2011,lepori2013}). \\
Conversely, approaching the line $\alpha =1$, the same multiplets tend to assume a 
continuous distribution whose decay becomes much slower. Even more interestingly, also when $|\mu|<1$
the even parity of the multiplets disappears, paralleling the presence of a unique ($Z_2$-even) ground state $|\mathrm{GS}\rangle$.  \\
\begin{figure}[h!]
\includegraphics[width=0.49\textwidth-2pt]{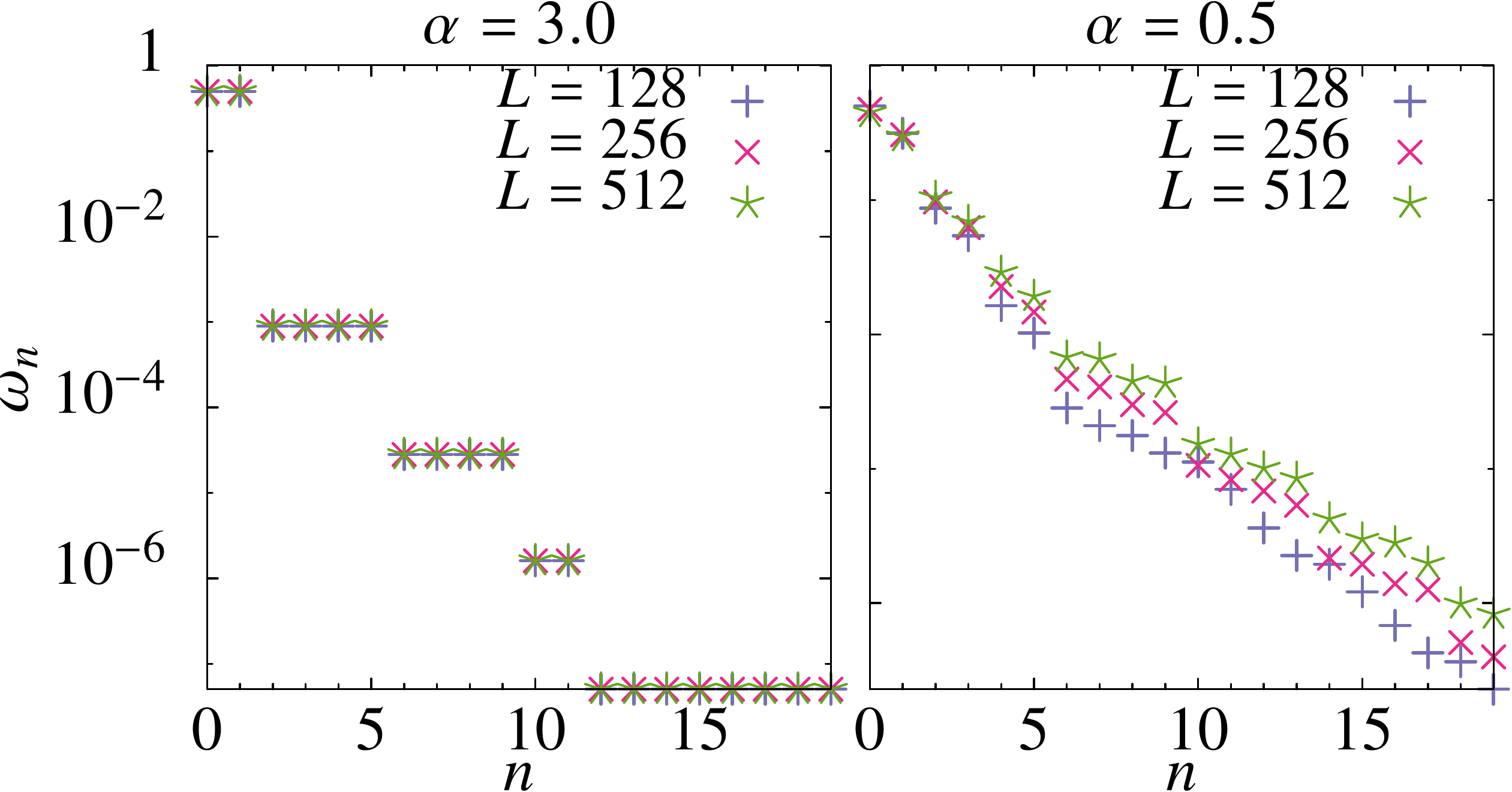}
\caption{Entanglement spectrum for the open LR paired chain in Eq.~\eqref{Ham} with $\mu= -0.5$, different $L$, and for $\alpha = 3$ (left panel) and $\alpha = 0.5$ (right panel).}  
\label{fig:ES}
\end{figure}
\begin{figure}[h!]
\includegraphics[width=0.485\textwidth-10pt]{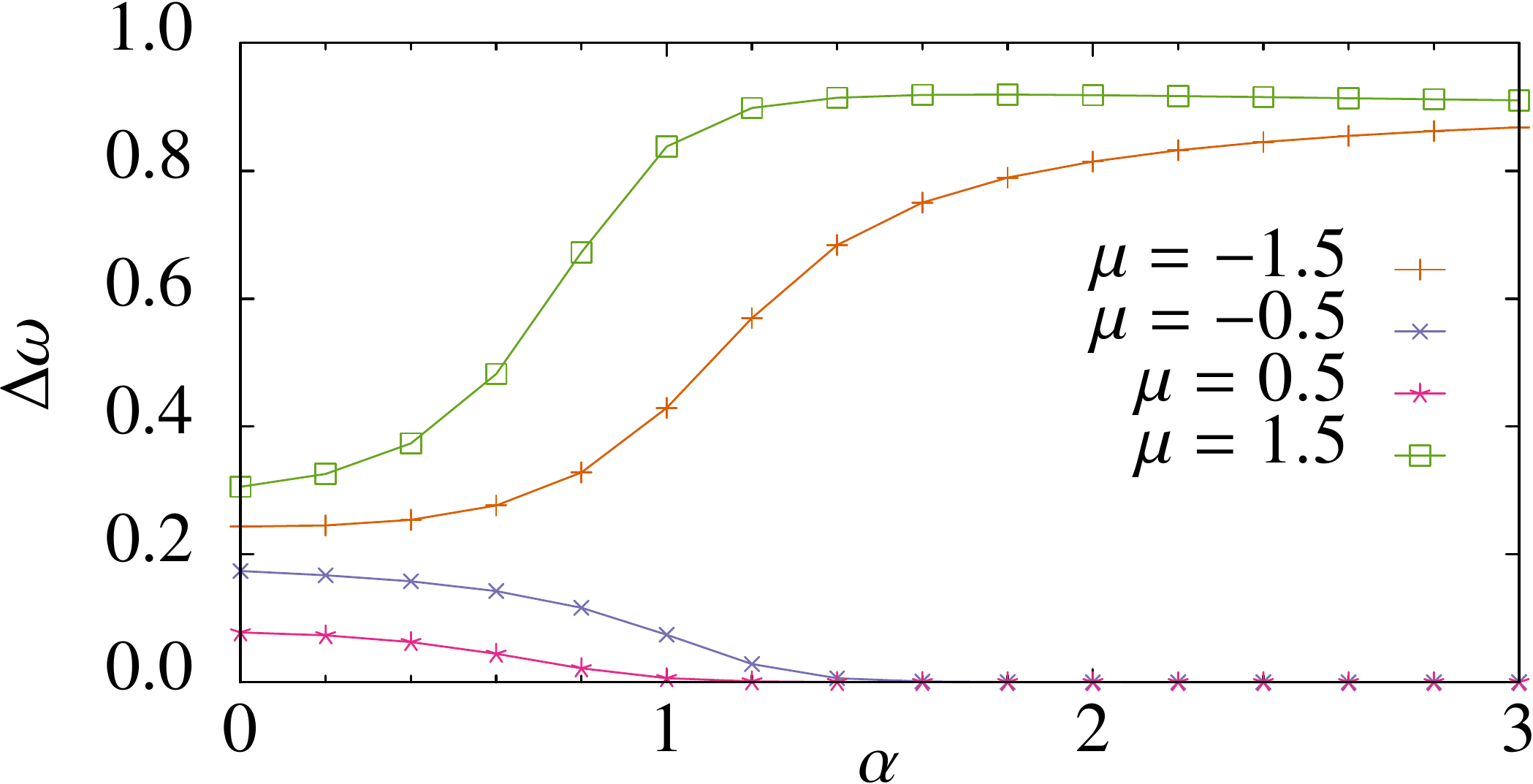}
\caption{Dependence on $\alpha$ of the difference $\Delta \omega$ between the two highest Schmidt eigenvalues (Schmidt gap)  for the open LR paired chain in Eq. (\ref{Ham}), different values of $\mu$ and $L = 512$. Notice that $\Delta \omega =0$ if $\alpha \gtrsim1$ and $|\mu|<1$, as expected in the SR limit.}  
\label{schgap}
\end{figure}
\begin{figure}
\includegraphics[width=0.45\textwidth-10pt]{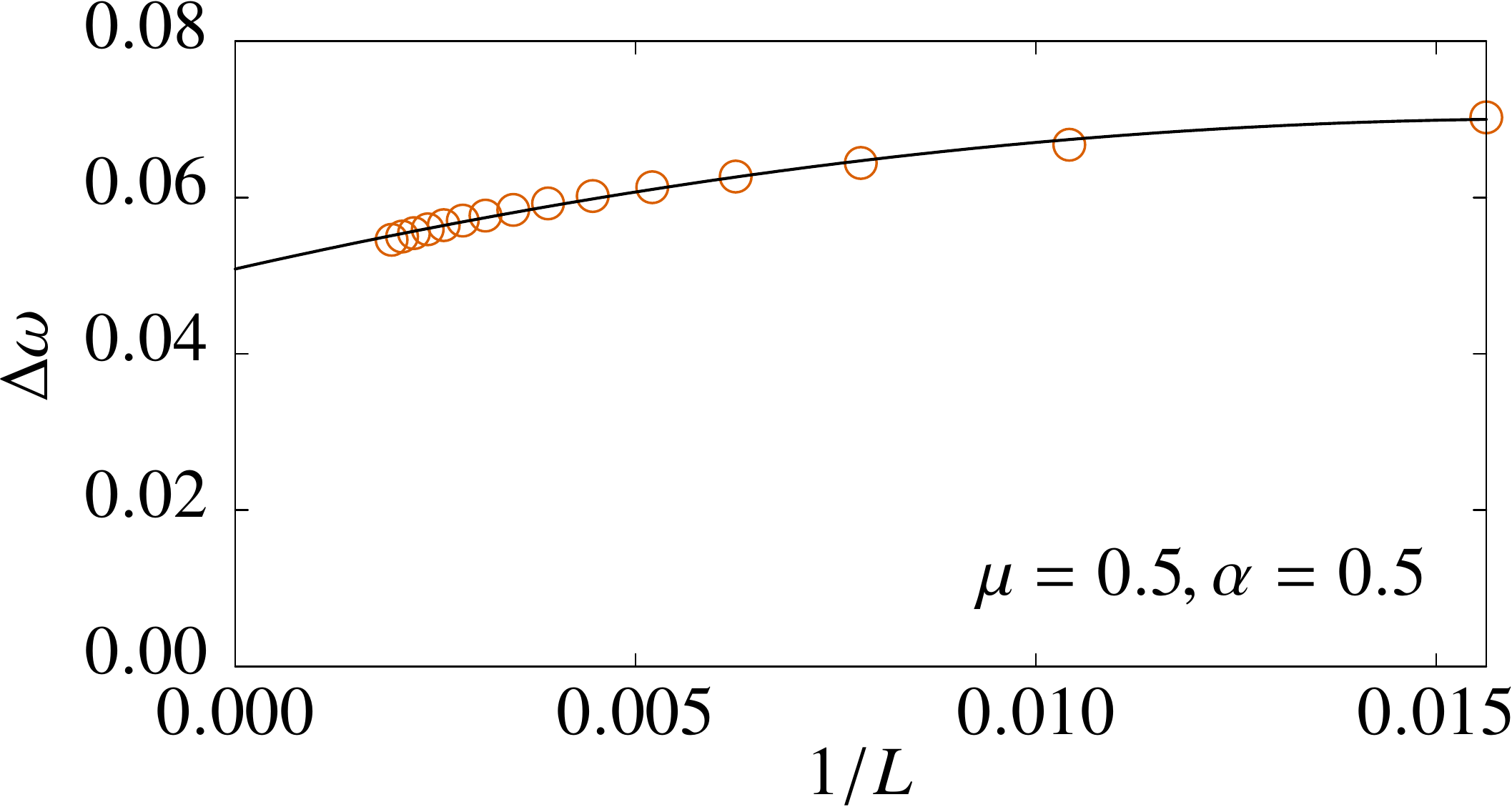}
\caption{Finite-size scaling of the Schmidt gap at $\mu = \alpha = 0.5$ (purple line in Fig. \ref{schgap}). The fitted value $\Delta \omega \approx 0.05$ strongly deviates from the much smaller values found at $\alpha >1$ and the same $\mu$ (see the main text). {\color{black}{We used a polynomial fit up to the fourth power in $\frac{1}{L}$.}}}
\label{schfs}
\end{figure}
The absence of constraints on the parity of the Schmidt multiplets is also shown in Fig. \ref{schgap}, where the behaviour of the difference between the  
two highest Schmidt eigenvalues, called "Schmidt gap'' \cite{dechiara2012}, is reported for different values of $\mu$ and $L = 512$. We see that, approximately below $\alpha = 1$, this quantity becomes nonvanishing for every $\mu$.\\
 In order to exclude that the nonzero values for the Schmidt gap below $\alpha = 1$ found in Fig. \ref{schgap} are due to finite-size corrections, we show in Fig. \ref{schfs} a finite-size scaling of the most unfavourable case in the former Figure  ($\mu = 0.5$), done with the data for chains with lengths from $L = 60$ to $L= 512$. This scaling yields at $\alpha = 0.5$ a value $ \Delta \omega \approx 0.05$ for the Schmidt gap, notably not far from the value at $L = 60$;  this fact indicates the limited role of the finite-size effects for the $\Delta \omega$. Notice finally that $ \Delta \omega \approx 0.05$ is much bigger than the value at $\alpha = 2$, where we obtain $\Delta \omega < 10^{-5}$. \\

The differences in the distributions of the ES in the range $|\mu|<1$ and on the two sides of the line $\alpha = 1$ confirms the appearance of purely LR phases below this threshold.
Even more interestingly, the present results stress once more the deep difference between the latter phases and the SR ones.\\
 Indeed it is known \cite{kitaev2010,turner2011}  that,  no matter the presence or the absence of interaction, only two phases (ordered and disordered) can be realized on a single SR Kitaev chain \cite{notetwophases}. The disordered phase of this model, having a single ground-state $|\mathrm{GS}\rangle$, displays no constraint on the number of Schmidt eigenvalues in each multiplet, so that multiplets with odd numbers of eigenvalues are also present.  In particular, the minimum degeneracy for a multiplet is 1. On the contrary, the ordered phase is characterized by even Schmidt multiplets. In particular, the minimum degeneracy for  a multiplet is 2.\\
The described  SR picture is not realized instead in the LR phases at $\alpha<1$. Indeed these phases display also single degeneracies in the Schmidt multiplets but, as inferred in Sections \ref{secbulk} and \ref{class}, they are disconnected from the disordered phase of the SR Kitaev chain.\\
This peculiar behaviour for the ES parallels the violation of the area-law and the appearance of massive edge states (when $\mu<1$), as we will discuss in more detail
in Section \ref{failES}, where formal reasons for the inapplicability of the arguments reported in \cite{turner2011} is also analyzed.

\section{Formal origin of the deviations from TWC}
\label{origin}

In this Section {\color{black} we investigate, at a more formal level,} the origin of deviations
from the TWC that can occur in LR quantum systems, analyzing 
the hypothesis at the bottom of the TWC and their possible {\color{black} inapplicability} in the presence of LR Hamiltonian terms.
The same analysis suggests that in general only some singularities in the energy spectrum can induce LR phases, while others preserve the SR phase content and in general the TWC.\\
We remember that, even if we are still dealing with superconducting phases, the main results of our discussion
are not restricted to this set of systems. Moreover no limitations are implied on the dimensionality of the considered LR quadratic fermionic models.

A key to understand the TWC {\color{black} in any dimension is based on the  classification of the topologically inequivalent continuous maps from the space of the lattice momenta $\b{k} \in [0, 2 \pi )$ (assumed to be a good quantum number, due to translational invariance in periodic systems) \cite{notehom}}, to a suitable grassmanian manifold $F$, induced by the matrix Hamiltonian $H(\b{k})$  (\cite{zirnbauer1996,zirnbauer1997,ludwig2008,kitaev2009,ludwig2009,ludwig2010}, \cite{book_intro} and references therein). {\color{black} These maps are defined univocally by some (sets of) integer numbers, generally called  winding numbers.}
In general  $F$ has the form of a coset space $G/\mathcal{H}$, being $G$ and $\mathcal{H}$ some groups.
In the absence of further symmetries, these manifolds are strongly constrained by the discrete anti-unitary charge-conjugation and time-reversal symmetries. \\
{\color{black} For instance,} in the particular case of spinless superconductors, as for the generalized Kitaev chains in Eqs. (\ref{Ham}) and \eqref{Ham2},  it is useful to classify the windings of the unit vectors  $\hat{n}_{\b{k}} = \frac{\b{n}_{\b{k}}}{|\b{n}_{\b{k}}|}$ such that $H(\b{k})$ can be written as  $H(\b{k}) = |\b{n}_{\b{k}}| \, \hat{n}_{\b{k}} \cdot \vec{\sigma}$, where $\sigma_i$ are the Pauli matrices (see \cite{tewari2012} for their differential expressions in the one-dimensional D/BDI classes).\\
However, as discussed in the previous Sections,  in the presence of LR Hamiltonian terms in real space,  singularities for $H(\b{k})$ and for its spectrum $\lambda(\b{k})$ can occur.
The behaviour of these divergences, say at a momentum $\b{k}_0$, strongly affects the definition of the {\color{black} winding numbers $w$ (defined for the one dimensional case in Section \ref{secbulk})} and the possible emergence of the LR phases. Indeed if 
$H(\b{k}_0 + \b{\epsilon})$ {\color{black} ($\b{\epsilon}$ being a infinitesimal quantity defining an open set around $\b{k}_0$)} does not depend explicitly on $\b{\epsilon}$ (in one dimension if  
$H(k_0+ \epsilon) = H(k_0 -\epsilon)$\big), the singularities at $\b{k}_0$ do not really spoil this definition {\color{black}{(in the particular case above} they are regularized dividing $H(\b{k})$ 
by  $|\b{n}_{\b{k}}|$: $\tilde{H}(\b{k}) \equiv \frac{H(\b{k})}{|\b{n}_{\b{k}}|}$\Big)},
as well as of the related winding numbers. This is the case for $g_{\alpha} (k)$ in Eq. \eqref{defg}. In the following we will quote these divergencies as \emph{first type divergences/singularities}. {\color{black} This set also includes, as a special case, the situation where only (say) $H(k_0+ \epsilon)$ diverges as $\epsilon \to 0$, while $H(k_0 - \epsilon)$ tends to a finite value; an explicit example of this situation has been proposed recently in \cite{bettles2017}.}\\
 On the contrary, when $H(\b{k}_0 + \b{\epsilon})$ depends explicitly on $\b{\epsilon}$ (in one dimension if $H(k_0+ \epsilon)  \neq  H(k_0 - \epsilon) \big) $, the path in the $H(\b{k})$ manifold experiences  not re-absorbable {\color{black} discontinuities}. For this reason, the {\color{black} winding numbers} of $H(\b{k})$
 that can define and classify the topology of the SR phases in every class of the TWC  are now ill defined \cite{notatop}.  In this condition the TWC can generally fail and new (LR)  phases can occur.  In the following we will quote these divergencies as \emph{second type divergences/singularities}. The different {\color{black} played by} the two types of divergencies will
 be probed explicitly in Section \ref{class}.
 
Notably in \cite{paper1}, using an effective theory close to the critical lines (where conformal invariance is explicitly broken), it has been shown explicitly for the model in Eq. \eqref{Ham} that divergences in $H(\b{k})$ of the second type  induce directly the violation of the area-law for the Von Neumann entropy. {\color{black} The same holds also for the phase transition at $\alpha = 1$ \cite{pezze2017}.} Based on the discussion above, we are lead to think that this parallelisms holds generally in gapped LR fermionic systems.

{\color{black}{Recently some attempts to define a topology for some LR fermionic systems appeared \cite{delgado2015}, exploiting the differential invariants in \cite{tewari2012}.
For the system in Eq. \eqref{Ham}} these attempts led to mathematically inconsistent results for $\alpha<1$:} indeed semi-integer winding numbers have been obtained in this condition, 
in spite of the fact that, being measured on closed $d$-dimensional loops, winding numbers should assume only integer values \cite{volovik,nakahara}. In this way the topology defined in terms of these winding numbers is mathematically not even well defined, as well as a connection between these numbers, calculated in the bulk, and possible edge excitations (see more details in SectionS \ref{class} and \ref{failES}).
However, in the light of our discussion, the mere appearance of winding numbers with fake semi-integer numbers (the same numbers instead well defined with integer values in the SR limit) can be interpreted as a clear physical diagnostic of new LR phases beyond TWC. This point will be discussed in better detail in Section \ref{class}.

The analysis above suggests that no new phase is expected in the presence of other singularities 
in higher-order derivatives of the spectrum $\lambda(k)$, as  for the Hamiltonians in Eqs. (\ref{Ham}) and Eqs. (\ref{Ham2})
in $k =\pi$ at every finite $\alpha >1$ (on the contrary, the same singularities has been found responsible of other LR effects, as
explained in the previous Sections).
In this way, a particular care is required for evaluating the regime {\color{black} $1<\alpha<\frac{3}{2}$ for the Hamiltonian in Eq. \eqref{Ham} ($\alpha_c = \frac{3}{2}$ being the critical value for $\alpha$ where the group velocity for the Bogoliubov quasiparticles diverges if $\mu \neq 1$), suspected in \cite{delgado2015} to have LR nature: massive edge states and fractional winding numbers.} For more details, see Sections \ref{class}, \ref{correlation} and  \ref{failES}.\\

We discussed above that in the SR limit topology can be encoded in some (sets of) {\color{black} windings number(s)}
induced by the mapping $\b{k} \to H (\b{k})$ itself \cite{ludwig2008,ludwig2009}. 
Exploiting a non-linear $\sigma$-model description of these grassmanian manifolds, in the past literature the TWC has been obtain directly \cite{zirnbauer1996,zirnbauer1997,ludwig2008,ludwig2009}.
Indeed $F$ is strongly constrained by the (anti-unitary) symmetries of the system under consideration, setting its topology class.
Remarkably, the same approach implicitly addresses the stability of the phases of the SR topological insulators {\color{black} and superconductors} against the 
introduction of a onsite disorder; indeed the latter ingredient is explicitly assumed and encoded.

{\color{black} In the presence of LR singularities in $H (\b{k})$, we can show that 
the non-linear $\sigma$-model construction, which leads to the TWC, cannot be performed, at least in the way derived following the 
standard approach \cite{wegner,efetov,belitz,fabrizio,dellanna2006,mirlin2008}, recalled in the Appendix \ref{bulk2} 
(where the proof of the inapplicability of the standard construction is provided for the first time). 
More in detail, the same approach generally yields 
a low-energy effective euclidean action whose kinetic part has the following form: 
\begin{equation}
S_{\sigma}[Q] \sim C \,  {\Tr}
(\b{\nabla} Q \, \b{\nabla} Q) \, ,
\label{expant}
\end{equation}
being $Q$ an effective matrix field and $C$ a constant. However, as shown in the Appendix \ref{bulk2}, 
the latter constant turns to be divergent in the LR phases, 
testifying the inapplicability of the $\sigma$-model construction, valid instead in the SR limit.} 

We finally comment that in LR systems the effect of disorder could be expected to be more 
dramatic than in SR models, spoiling the divergences in the energy spectrum that originate the LR phases and all the 
other LR peculiarities, however this possibility will be ruled out in Section \ref{dissec}.

\section{Towards a bulk classification of LR phases}
\label{class}

In this Section we deal with the problem of classifying the purely LR phases of quadratic fermionic Hamiltonians, 
{\color{black} also defining a nontrivial topology for them.} We perform the discussion for one-dimensional systems  at first.

We discussed in Section \ref{origin} that {\color{black} winding numbers}
for the matrix Hamiltonians $H(k)$ are apparently not useful, being ill defined in the LR phases. The reason for that inapplicability lies on the discontinuities encountered in the path on the $H(k)$ manifold as $k$ varies in the Brillouin zone,
e.g. in the correspondence of second type divergences (say at $k_0$), where $H(k_0+ \epsilon)  \neq  H(k_0 - \epsilon)$.\\
{\color{black} Another approach, still connected with the first one and also valid in the SR limit,} is to 
consider  the Berry phase  \cite{berry1984}
\beq \
\Phi  =  i \, \int_{BZ} \mathrm{d} k \,  \langle u_{k} | \partial_{k}  u_{k} \rangle \, ,
\label{berryphase}
\eeq 
gathered again as $k$ varies along the Brillouin zone. The vector $| u_{k} \rangle$ is an eigenvector of $H(k)$ 
and the integral extends on the Brillouin zone.
The same approach has been exploited in \cite{delgado2015}  for the particular Hamiltonian in Eq. \eqref{Ham}.\\
For sake of generalization and in order to identify  ambiguity problems in the definition of a nontrivial LR topology, it is useful to discuss here the main step of the calculation for the models in Eqs. \eqref{Ham} and \eqref{Ham2}, both in the SR limit and in the LR one. In both of the regimes the same calculation proceeds in a very similar way.\\
We notice at first that, since $ \langle u_{k} | u_{k} \rangle = 1$, we have that,  if we fix $| u_{k} \rangle$ real, $\langle u_{k} | \partial_{k}  u_{k} \rangle = 0$, unless $|u_{k} \rangle$ and/or $| \partial_{k}  u_{k} \rangle$ are singular. In all the cases considered in the present paper $|u_{k} \rangle$ are well defined (finite), as well as the Bogoliubov transformations leading to them \cite{nostro,paper1}, while the second possibility can be realized, being $|u_{k} \rangle$ discontinuous. This fact holds in the correspondence of second type singularities (say again at $k_0$):
\beq
| u_{k} \rangle = | v_{1 \, k} \rangle +  \theta(k-k_0) | v_{2 \,  k} \rangle \, ,
\eeq        
or, equivalently, 
\beq
| u_{k} \rangle = M_{k_0}(k) \,  | v_{k} \rangle = \big({\bf I} + \theta(k-k_0) \, N(k)\big)  \,  | v_{k} \rangle  \, ,
\label{trasf2}
\eeq 
{\color{black} being $M_{k_0}(k)$ and $N(k)$ suitable  matrices.}
Notice that $N(k)$ is continuous through $k_0$: $N(k_0 + \epsilon) = N(k_0 - \epsilon) \equiv N(k_0)$, moreover we have continuity in $k_0$ for  the energy  $\lambda(k)$ of  $ | u_{k} \rangle$:  $\lambda(k_0 + \epsilon) = \lambda(k_0 - \epsilon) =\lambda(k_0) $. The latter fact is central to assure the Berry phase to be well defined. \\
The matrix $M_{k_0} (k)$
 transforms locally $H(k)$ where the singular point $k_0$ is encountered:
\beq
H(k_0+ \epsilon)  =  M_{k_0}(k_0 + \epsilon) H(k_0 - \epsilon) M_{k_0}^{-1}(k_0 + \epsilon) \, 
\label{glue2}
\eeq
so to assure that
\beq
\lambda (k)  =  \langle u_{k}|H(k)|u_{k} \rangle
\eeq
varies continuously passing through $k_0$. We stress that, in spite of the matrix $M(k)$,
the nature of the Berry phase $\Phi$ is purely abelian in all the cases analyzed in this paper, since no degeneracy for the ground state occurs.

It is straightforward to show that, passing through $k_0$ from below, a Berry phase 
\beq
\Phi_{k_0} = - \frac{\pi}{4} \,  \langle v_{k_0}|M(k_0)^{-1} \, \partial_k M(k)_{| k_0}|v_{k_0} \rangle
\eeq
is gathered. This expression can be easily evaluated
writing $\theta(k-k_0) = \frac{1}{2} \, \big(1+ \mathrm{sign}(k-k_0)\big)$ and using the complex expression for the derivative of the $\mathrm{sign} (k)$ function:
\beq
 \frac{\partial \, \mathrm{sign} (k)}{\partial k} = i \, \pi \, \delta(k) \,  \mathrm{sign} (k) \, .
\eeq

Importantly, the calculation scheme described above works completely equivalent for the discontinuities occurring in SR systems and 
for the LR ones from the second type singularities.
Exploiting the same scheme, it is easy to show that:
\begin{itemize}

 \item for the Hamiltonian in Eq. \eqref{Ham} ($\alpha < \infty$, $\beta = \infty$), $ |u_k\rangle = (\cos \theta_k, \sin \theta_k)$, with 
 \beq
 \theta_k = - \frac{1}{2} \, \arctan \Bigg(\frac{f_{\alpha}(k)}{(\mu + {\color{black}w}\,g_{\beta = \infty} (k))} \Bigg) \, ,
 \label{theta}
 \eeq   
 and $g_{\infty} (k) = \cos k$.
 Two discontinuities in $\sin \theta_k$ in  $|u_k \rangle$ arise \big(at $k_{1,2}(\alpha)$\big) if $|\mu|<1$ and for every $\alpha$,  because there the diagonal 
 terms  in Eq. \eqref{tbHam} change sign. Passing through each of them from below, a $\frac{\pi}{2}$ phase is gathered. Indeed if  $\alpha >1 $ they give rise to the total Berry phase $\Phi = \pi$  proper of the phase with massless edge modes. In these cases
the matrix $M(k)$ in Eq. \eqref{trasf2} around $k_{1,2}(\alpha)$ reads: $M(k) = \big({\bf I} + \theta (k- \pi) (\sigma_x - {\bf I}) \big)$.\\
 Another discontinuity in $\sin \theta_k$ appears if $\alpha<1$, because of a 
 second type singularity
 at $k = \pi$, responsible of the outcome of LR phases. More in detail, this is due to the 
 behaviour of $f_{\alpha} (k+\pi)$: {\color{black}$\lim_{k \to \pi^+}f(k+\pi)=+\infty$, and $\lim_{k \to \pi^-}f(k+\pi) = - \infty$.}
 We find that, passing through $k = \pi$ from below, a $\frac{\pi}{2}$ phase is gathered if $\mu >1$, while 
 a $-\frac{\pi}{2}$ phase is gathered if $\mu <1$. In these cases
the matrix $M(k )$ in Eq. \eqref{trasf2} reads, around $k = \pi$, $M(k) = \mathrm{sign} (\mu-1) \,  \mathrm{diag} \big(1, \mathrm{sign} (\pi - k)\big)$, the same found in \cite{delgado2015}.
 
  Collecting all these partial results, 
 it is found that $\Phi = 0$ if $|\mu| >1$
 and $\Phi = \pi$ if $|\mu| <1$ when $\alpha>1$,
while at $\alpha<1$ we obtain $\Phi = - \frac{\pi}{2} = \frac{3 \pi}{2}$ if $\mu <1$ and $\Phi =  \frac{\pi}{2}$ if $\mu >1$.  

These findings show the presence of two purely LR phases at $\alpha <1$ disconnected from the SR ones (having different values of $\Phi$), as described 
in the Section \ref{kitaev}. Moreover, they indicate a nontrivial topology for (at least) one of them, see more detail in the following of the Section.

\item for the Hamiltonian in Eq. \eqref{Ham2} with $\alpha = \infty$ and $\beta < \infty$ (LR hopping, SR pairing),
exploiting the expression for $\theta_k$ in Eq. \eqref{theta}, only phases with $\Phi = 0$ and $\Phi = \pi$ are found, also at $\alpha <1$. 
Correspondingly, a qualitatively equal situation as for the Hamiltonian in Eq. \eqref{Ham} at $\alpha>1$ takes place: massless edge modes are found if $\Phi = \pi$, while no edge mode at all if $\Phi = 0$. These results match our expectation that first type singularities, as for $g_{\beta} (k)$ ($f_{\infty} (k) = \sin k$ is regular), do not induce alone LR phases. Indeed these singularities yield $M(k) = {\bf I}$.

\item  for the Hamiltonian in Eq. \eqref{Ham2} with both finite $\alpha$ and  $\beta$, we obtain:

\emph{a)} if $\alpha < \beta$, below $\alpha = 1$ 
we find, as $\mu$ varies, zones with $\Phi = \frac{\pi}{2}$ and $\Phi = \frac{3 \, \pi}{2}$, as 
for the case $\alpha < \infty$ and $\beta = \infty$ (Hamiltonian in Eq. \eqref{Ham}).
In particular, at fixed $\alpha$ the second zone occurs at smaller $\mu$ compared with the first one.
Correspondingly,  the same content of massive edge states is found. {\color{black} The quantization of the Berry phase,
required to assure topological stability,
will be discuss at the end of the present Section;}

\emph{b)} if  $\alpha > \beta$, below $\alpha = 1$ we find, as $\mu$ varies,  zones with $\Phi = \pi$ and $\Phi = 0$, as 
for the Hamiltonian in Eq. \eqref{Ham2} with LR hopping only. Indeed here the contribution of the (first and second type) singularities at $k = \pi$ from {\color{black}$g_{\beta}(k+\pi)$ and $f_{\alpha}(k+\pi)$} effectively cancel each other.  At fixed $\alpha$, the first zone occurs again at smaller $\mu$ compared with the second one.  If $\Phi = \pi$  massless edge modes are found, while no edge mode at all when $\Phi = 0$. 

\emph{c)} if $\alpha = \beta$, at $\alpha < 1$ we find, as $\mu$ varies, zones with $\alpha$ dependent Berry phases: $\Phi = - \pi \, K(\alpha)^2$ and $\Phi = \pi\, \Big(1- K(\alpha)^2 \Big)$,
with
$K(\alpha) = - \sin\Big(\frac{1}{2} \, \arctan \frac{1}{\tan (\frac{\pi}{2} \, \alpha )} \Big)$. Again at fixed $\alpha$, the second zone occurs at smaller $\mu$ compared to the first one, in this regime massive edge states have been previously  found, for the first time \cite{paperdouble}. The contribution $\propto K(\alpha)$, due to the second type singularity,
vanishes at $\alpha = 1$, as expected, while it tends to $-\frac{\pi}{\sqrt{2}}$ at $\alpha = 0$ (then the same values for $\Phi$ as in the first 
example above are recovered). Strikingly, the quantity $\Phi$ varies continuously in the range $0 \leq \alpha \leq 1$, so that apparently it cannot be assumed \emph{a priori} as an order parameter to distinguish SR and LR phases (while it appears effective to discriminate between the LR phases) {\color{black} and to assure their topological stability against perturbations. However, a way to remove this problem will be discussed close to the end of the present Section}. Moreover, the presence 
of these phases can be proven also by the ground-state degeneracy arguments in Section \ref{phased}.

\end{itemize}

We notice that, as resulting from the discussed examples, 
 differently from the SR systems, for LR ones the appearance of a nonzero Berry phases, 
 does not imply the presence of edge states in general. 
For instance, in the first example (LR pairing only), in the regime $\mu >1$  a phase  $\Phi = \frac{\pi}{2}$ is derived, {\color{black} which however does not correspond to the presence of massive edge states}, in spite of the fact that this value is different from $\Phi = 0$, 
{\color{black} the proper value in the absence of edges modes (as the empty space beyond the edges themselves)}. The difference stems from the second type divergence at $k = \pi$; this contribution is present for \emph{all} the LR phases at every $\mu$, indeed is exactly the one discriminating SR and LR phases.\\
The latter example indicates the necessity for a more specific criterium to link the Berry phase  $\Phi$ with the possible presence of edge states and with their properties in LR phases. 
From all the analyzed examples, {\color{black} we are led to think} that, given a certain model having different LR phases with Berry phases $\{\Phi_i\}$, 
edge states occur {\color{black} whenever $\tilde{\Phi}_i = \Phi_i - \Phi_M \neq 0$, where $\Phi_M$ is the common contribution present in all the LR phases,  not quantized in general (as exemplified in  the case \emph{c)} above), and only discriminating them from the SR phases (indeed derives directly from the \emph{second type} singularities in the quasiparticle spectrum).  Then $\Phi_M$ defines the trivial LR topology. Consequently, the quantized quantity $\tilde{\Phi} \neq 0$ defines instead the \emph{nontrivial} LR topology,  related with the massive edge states.
After (and only after)  that the LR contribute $\Phi_M$ has been properly identified, the described subtraction procedure amounts to avoid the LR singularities in the calculation of the Berry phase accordingly to Eq. \eqref{berryphase}.}

The Berry phase approach, with the caveats discussed above,  looks suitable for extension to classify LR phases of (at least) one-dimensional Hamiltonians with different symmetry content from the BDI class examined in the present paper. \\

{\color{black} The same ambiguity encountered for the Berry phase is found in the attempt to define a LR topology by the winding number $w$, defined as in Section \ref{secbulk}.  The two approaches are linked together by a classical result by Berry \cite{berry1984}: for the particular case of a $2\times 2$ real matrix Hamiltonian, as in Eq. \eqref{tbHam}, it holds
\beq
\Phi  = \pi \, w \, .
\label{twoinv}
\eeq
The mentioned ambiguity is due again} to the primary difficulty of identifying the \emph{trivial} LR topology. For instance, both for the Kitaev Hamiltonian  with LR pairing only and for the Hamiltonian 
 with also LR hopping, the path of the vector $\b{n}_k$ (defined in Section \ref{origin}) in the two LR phases, as $k$ changes from $0$ to $2 \pi$, 
is a semi-circle around the origin $(0,0)$, {\color{black} which is closed by a jump between the two ending points of the semi-circle,} a direct consequence of the LR divergences.  {\color{black}Notably, the closed paths (considering also the jumps) in the two phases differ by an entire circle, as noticed in \cite{delgado2015}. This fact allows us to conclude that the two phases have different topologies ($\Delta w = |w_1-w_2| =1$, indeed $w$ counts the number, also not integer in the LR regime, of closed loops around the origin) and to suspect the appearance of massive edge states in one of them. \\
In order to discriminate between the LR phase with trivial topology from other ones with nontrivial topology and massive edge states we can use the relation 
in Eq.~\eqref{twoinv}, leading back to the same discussion for the Berry phase: identified as trivial the quantum LR phase having $\Phi_M$ and winding $w_1$ (generally not integer), the LR topological phases are characterized by the integer numbers $\Delta w$ and by the related phases 
 \beq
\tilde{\Phi}  = \Phi- \Phi_M= \pi \, \Delta w \, .
\label{twoinv2}
\eeq
 Finally, Eq. \eqref{twoinv2} ensures the quantization of the Berry phase also in the LR regime and its consequent stability to perturbations.\\}

In conclusion, the discussion of this Section {\color{black} suggests that the Berry phase and the generalized winding number can be still useful} to define a nontrivial topology in LR quantum systems, {\color{black} resulting in the presence of massive edge states,} provided the proper primary identification of the \emph{trivial} topology.  
This possibility {\color{black} should be valid also for} LR quantum systems with higher dimensionality (for a review on the same methods applied to general SR topological insulators  see for instance \cite{book_intro} and references therein), indeed no additional obstructions seem to appear in these conditions. 
 The evaluation of the two approaches on specific higher dimensional cases (an issue also involving the problem of defining LR topological numbers entirely in terms of local quantities/currents neglecting path discontinuities) deserves deep future attention in our opinion.

\section{Partial failure of edge characterization: weakening of bulk-boundary correspondence}
\label{failES}

In Sections \ref{kitaev} and \ref{ES} we explained that the distribution of the ES in the LR phases of the Hamiltonian in Eq. \eqref{Ham} below $\alpha = 1$ does not insert in the SR classification scheme derived in \cite{turner2011} for the one-dimensional BDI symmetry class.\\
 In this Section we {\color{black} investigate at a more formal level} the origin of this deviation. This analysis will yield further information on other LR peculiarities, for instance the nature of the massive edge states, their link with the bulk excitations and the asymptotic behaviour of correlation functions. Moreover the same analysis will appear suitable for almost straightforward extensions to other one-dimensional symmetry classes.

\subsection{On the behaviour of the LR correlation length}
\label{correlation}

The discussions of the previous Sections require further investigations on the definition of the correlation length $\xi$ in LR systems.\\
In gapped SR systems the correlation length can be defined in various ways. The most common one is by the asymptotical exponential decay of the two-points correlation functions. Let us take a field $\phi(x)$ of a generic model, the correlation function we will consider is the following 
\beq
C(x) = \langle GS | \phi (x) \, \phi(0) | GS \rangle_{x \to \infty}  \sim e^{-\frac{x}{\xi}}  \,.
\label{defxi}
\eeq
This quantity is expected, instead, to diverge at the continuous critical points, where the correlation functions decay algebraically. Moreover, 
at criticality, violation of the area-law for the Von Neumann entropy occurs \cite{wilczek,calabrese}, a fact which is at the base of the scaling 
hypothesis and of the effective RG description for critical phenomena (see e.g. \cite{vicari2002}).

For a $(1+1)$-dimensional SR quantum system, a second definition for $\xi$ can be given close to (but not exactly at) the a critical point, 
by the asymptotic
scaling of the Von Neumann entropy. Given $l$ the length of a subsystem, the entropy goes 
\beq
S(l) \sim  \frac{c}{6} \,  \mathrm{ln} \, \xi  \quad \mathrm{as} \quad l \to \infty \, ,
\label{xiVN}
\eeq 
$c$ denoting again the central charge describing the critical point and $S(l)$ being defined as in Section \ref{phased} \cite{calabrese}. This law is similar to the one in Eq. \eqref{dev}. 
Equation \eqref{xiVN} implies that the Von Neumann entropy saturates for large $l$, to a \emph{finite} value, function of $\xi$. Importantly the latter definition only relies on the existence of a critical point, described by a conformal theory. For this reason, although the two definitions for $\xi$ in Eqs. \eqref{defxi} and \eqref{xiVN} match each others for SR systems (up to constants of order 1), the second one appears more suitable for generalizations to LR models, at least until conformal invariance is preserved.\\
Applying the first definition for $\xi$ in Eq. \eqref{defxi} to the LR Kitaev chains (as well as to the notable examples in \cite{deng2005,koffel2012,paperdouble}), the hybrid decay for correlation functions mentioned in the Introduction yields, at every finite $\alpha$, an infinite correlation length, $\xi \to \infty$.
However, this result does not match the fact that these systems display, up to critical (model dependent) $\alpha$, both a saturation of the Von Neumann and conformal points, such that Eq. \eqref{xiVN} is valid and $\xi$ is expected to be finite. Moreover, in the particular cases of LR Kitaev chains in Section \ref{kitaev}, if $1<\alpha <\infty$, the realized phases are continuously connected with the ones in the SR limit, so that a divergence of $\xi$, according to the definition in Eq. \eqref{defxi}, looks definitively odd.

For these reasons, a unique definition $\xi$ for LR systems appears an open issue. 
 At least for the models that we are analyzing, a partial help comes from the two points correlation functions $g_1 (R) \equiv \langle a^\dag_R a_0 \rangle$ and 
$g_1^{\text{(anom)}}(R) \equiv \langle a^\dag_R a^\dag_0\rangle$. Indeed their hybrid decay at finite 
$\alpha$ is characterized by {\color{black} a typical distance $R^*$, increasing with $\alpha$, and separating the exponential and the algebraic decay regimes, at short and large separations respectively \cite{deng2005,koffel2012,nostro,paperdouble}.
 In particular, the exponential part becomes practically absent at $\alpha \lesssim 1$, so that the algebraic tail strongly dominates.

In the light of this behaviour, in the large $\alpha$ limit (where Eq. \eqref{xiVN} holds and the algebraic part of the correlations also begins at very large separations and with very small magnitudes), a typical length for correlations can be still defined by the exponential decay close to $R^* \to \infty$. \\}
 On the contrary, in the regime $\alpha \lesssim 1$, where the exponentially decaying part is negligible, a characteristic length is not available any longer. In this regime the system effectively behaves like a SR system at criticality, as one can understand from at the area law violation for the Von Neumann entropy and from the continuous distribution for the ES, described in Section \ref{ES}. 
The parallelism for the latter quantities is also someway justified by the fact that if $\alpha <1$ the correlation functions $g_1 (R)$ and 
$g_1^{\text{(anom)}}(R)$, which determine alone the ES and the Von Neumann entropy in quadratic fermionic models \cite{peschel2011}, 
decay algebraically with an exponent  $\gamma <1$ always \cite{nostro,paperdouble,paper1}, as for the critical SR quadratic Hamiltonians in one dimension (see \cite{muss} and references therein).  \\
This analogy naturally leads us to conjecture that $\xi$ effectively diverges for $\alpha \lesssim 1$. In the next Section we will check that this hypothesis is able to explain the emergence and various properties of the LR phases.  We leave as open and important issues the probe on other models and the rigorous justification of this hypothesis, as well as a definition for $\xi$ in the regime $\alpha \gtrsim1$.

\subsection{Inapplicability of the {\color{black} edge operators approach} for the ES}

 In this Section we analyze how the ES characterization for the one-dimensional BDI symmetry class discussed in \cite{turner2011} can be not applicable in the presence of LR Hamiltonian terms.

 In that paper the discussion is based on the analysis of certain edge operators $Q_{R/L}$, able to induce the bulk transformations belonging to the invariance group of the considered Hamiltonian, at least involving the states with highest Schmidt eigenvalues (then more likely after a bipartition). In this way, the (anti)-commutation relations between the operators $Q_{R/L}$ and with the generators of the symmetries of the Hamiltonian are able to constrain entirely the ES, classifying without ambiguities the SR fermionic phases.
 The properties of the edge operators reflect the ones owned by the edges, in particular due to the possible presence of localized modes on them. 
For one-dimensional quantum systems, both fermionic and bosonic (as spin models, 
analyzed in \cite{pollmann2010}), this construction formalizes the so called bulk-boundary correspondence conjecture. 
 
 The demonstration of these results starts, in \cite{turner2011}, showing first that local operations performed asymptotically far from the edges  cannot change the  (highest part of the) entanglement 
 content of the considered system. 
 
 A crucial property exploited at this first step is the cluster decomposition for correlation functions.
It is expected therefore that fermionic systems violating  the cluster decomposition can display
important deviations from the classification scheme in \cite{turner2011}.  However, this property is preserved for the 
Bogoliubov ground-state of the LR Kitaev chains in Eqs. \eqref{Ham} and \eqref{Ham2} \cite{nostro,paperdouble,paper1,notecluster}.

 The second step of the discussion in \cite{turner2011} is
 the explicit construction of the boundary operators, relying
on a MPS-like approach (valid for both  fermionic and spin models).  This construction is valid again for the highest Schmidt eigenstates and it requires the finiteness of the correlation length $\xi$.
In particular the error in the implementation of the Hamiltonian symmetry transformations on these states  by operators involving $l$ sites from the edges
scales as $\sim e^{-\frac{l}{\xi}}$.\\
However, for our models,  within the LR phases,
$\xi$ appears effectively divergent, as explained in Section \ref{correlation}.
Moreover, related with the divergence of $\xi$ and as required implicitly by the MPS-like approach used in \cite{turner2011}, 
the fulfillment of the area-law for the Von Neumann entropy in gapped regimes results a necessary condition for the validity 
of the edge operators construction: this ingredient is again not present for $\alpha <1$ in the LR systems studied in the present paper.
Notably a logarithmic violation of the area-law in a gapped regime
is already sufficient to determine 
important deviations from the SR picture, but even more dramatic deviations are expected in LR systems where the area-law 
is substituted by an almost volume-law, a situation described for instance in \cite{gori} and which deserves an 
further investigation in our opinion.

Finally, we comment that from the previous discussion it appears that the loss of validity of the cluster decomposition property implies 
the area-law violation; the opposite implication, instead, is not true in general
(as also exemplified by the SR critical systems): the area-law violation seems to indicate the loss of validity of the cluster decomposition in its \emph{exponential} form only \cite{kastoryano2013}.

\subsection{Nature of the massive edge states and weakened bulk-boundary correspondence}
\label{natedge}

The analysis of the last Subsection can help to shed light on the nature of the massive edge states found for the LR Hamiltonians 
 in Eqs. \eqref{Ham} and \eqref{Ham2} \cite{nostro,paperdouble,alecce2017}. \\
 Indeed the inapplicability of the discussion in \cite{turner2011}, based on the action of the edge
operators $Q_{R/L}$, suggests that, oppositely to the SR limit, the purely LR phases cannot be characterized entirely by their edge structure.
Indeed symmetry operations on a bulk state cannot be represented faithfully by operations near the two edges.
For the same reason, a certain bulk structure, for instance related to the ES, 
 does not reflect directly in the properties of the two edges (e.g. the presence of localized modes). 
In this sense we have a violation of the so-called bulk-boundary correspondence, at the base of the TWC in the SR limit.

The picture defined above seems not to match entirely with the discussion done in the previous Sections on the LR paired Kitaev chain. Indeed there 
the appearance of massive edge states (with mass $m$), below $\alpha = 1$ and for $\mu<1$, has been found to parallel a nonvanishing Berry phase calculated in the bulk and a consequent a nontrivial LR topology. However, in this situation, in spite of the double edge localization 
{\color{black}of the first Bogoliubov wavefunction} (with positive energy) $\ket{m}$, 
no real distinction between left and right edge modes can be made, oppositely to the SR limit.
Roughly speaking, below $\alpha = 1$ the two edges are so correlated each others and with the bulk, that a rigorous definition of localized modes on each of them, distinguished from the bulk dynamical excitations, is not allowed any longer, not even in the thermodynamic limit. Such an important correlation is testified by the algebraic decay tails of the edge wavefunctions, strongly dominating at $\alpha \lesssim 1$ \cite{nostro,paperdouble}. In turn, the relevant overlap of the latter tails {\color{black} can justify an hybridization mechanism of the SR Majorana modes \cite{paperdouble,neupert2016},} responsible for the appearance of the massive edge states $\ket{m}$, in analogy with the situation occurring  at finite sizes in the SR limit  \cite{kitaev}.\\
 The described situation, apparently peculiar of the LR quantum systems, can be quoted as \emph{weakened bulk-boundary correspondence}. {\color{black} More in detail, this definition denotes the situation where a nontrivial LR topology still reflects on the presence of states localized on the edges, but these states have a nonzero mass and consequently a dynamics not separable from the bulk one (in the sense that no modes localized on a single edge can be defined from these bulk states), as happens instead in the short-range limit.}

Exploiting the Bogoliubov construction of the bulk states and reviewing the standard construction of the edge modes above $\alpha = 1$ \cite{refpar}, in the Appendix \ref{appedge} we show the correctness of the picture described above. We argue in particular that the nonzero mass for $\ket{m}$ at $\alpha <1$ {\color{black} corresponds to 
the impossibility of defining}, from $\ket{m}$, two states localized separately on the left-hand and the right-hand edges. 
The same result implies in itself the inapplicability, for our LR models, of the edge characterization in \cite{turner2011} {\color{black} and of the bulk-boundary correspondence valid for SR systems}, in favour of the described weakened version.

We conclude the present Subsection noticing that the important correlations between edge and bulk dynamics, resulting in the nonvanishing mass $m$ for the analyzed one-dimensional examples, could indicate the absence of edge conductivity for LR topological insulators {\color{black} and superconductors} with dimensionality bigger than $1$, where long-range correlations {\color{black} (as in the $\alpha \to 0$ and mean field limits)} are even enhanced. In our opinion, it is highly worthy to probe this conjecture in future investigations.

{\color{black}
\section{Stability of the LR phases against local disorder}
\label{dissec}

In this Section we investigate the stability of the LR phases for $\alpha <1$ of the Hamiltonian in Eq. \eqref{Ham} against local disorder.  
The present specific study can be easily generalized to other LR non interacting models, also in higher dimensions.

We inferred in Section \ref{origin} the inapplicability of the $\sigma$-model construction, valid for SR systems, in LR free models at small enough $\alpha$, 
due to a type of divergence in their energy spectrum at some momenta.
The same construction encodes the effect of a local disorder and it allows to derive directly the TWC (see e.g. \cite{ludwig2009}).\\
This result is someway counterintuitive, since disorder could be expected 
instead to smear and/or localize the divergences in the spectrum (and also the ones in its higher order derivatives), 
spoiling all the LR features.\\
In this Subsection we infer that this possibility can be ruled out, at least for free one-dimensional LR models, like those in Eqs. \eqref{Ham} and \eqref{Ham2}, 
elaborating the results got in a previous study about the effect of local disorder on a spinless fermionic chain with LR hopping, performed in \cite{rodriguez2003}.\\
We assume to add to the Hamiltonian in Eq. \eqref{Ham} a onsite energy term $\sum_{i} \, \epsilon_i \, a_i^{\dagger} a_i$, $\epsilon_i$ being randomly distributed in an interval $\big[- \frac{\eta}{2}, \frac{\eta}{2} \big]$,  so that 
the standard deviation is $\sigma \propto \eta$.  \\
In momentum space the effect of the disorder Hamiltonian term $H_D$ is to mix the quasiparticles of the clean system ($H_D = 0$), possibly resulting 
in a localization of them in a restricted region of the entire system. However, this mechanism is efficient for the singular states only if the 
magnitude of the disorder $m_D \sim \frac{\sigma}{L^{1/2}} \sim \frac{\eta}{L^{1/2}}$ 
is at least comparable with the distance between the energy levels close to the singularity, $k\approx k_0$ 
(in our case $k\approx \pi$), 
$\delta \lambda \sim \frac{\eta}{L^{(\alpha-1)}}$, as shown rigorously in \cite{rodriguez2003}. 
From the two scaling laws for $m_D$ and $\delta \lambda$, we expect that, for $\alpha < \frac{3}{2}$, 
localization of the singular states does not occur, therefore the typical disorder 
as in Eq. \eqref{dis1} should not spoil the LR phases below $\alpha = 1$.\\
On the contrary, the states far from $k = k_0$, instead, 
can be generally localized by $H_D$ \cite{rodriguez2003}. For this reason the disorder 
can even highlight the role played by the singular states.
}

\section{Similar entanglement behaviour in the long-range Ising model}
\label{LRIsec}

In the previous Sections we argued that in gapped non-interacting fermionic systems 
{\color{black} the appearance of the area-law violation for the entropy and the peculiar behaviour of the ES signal new purely LR phases, 
induced by singular dynamics.} We would like to probe now this picture on other LR systems, 
for instance spin models or interacting fermionic systems. For this reason, in this Section 
we consider another paradigmatic LR system, the LR Ising model, recently studied both theoretically
\cite{koffel2012,paperdouble,lepori2016} and experimentally \cite{exp1,exp2}. {\color{black} After recalling, in Subsection \ref{isa}, the main features of the model, in Subsection \ref{isb} we discuss the 
behaviour of the ES, focusing on the regime $\alpha<1$, which is the main result of the present Section.}  

\subsection{Phase diagram and ground-state properties}
\label{isa}

The Hamiltonian of the LR Ising antiferromagnetic chain reads:
\begin{equation}
H_\text{LRI} = \sin \theta \, \sum_{i = 1}^{L-1} \sum_{\ell =1}^{L-i} \frac{1}{\ell^\alpha} \, \sigma^{(x)}_{i}\sigma^{(x)}_{i+\ell} + \cos \theta \, \sum_{i = 1}^{L} \sigma^{(z)}_i \, .
\label{LRI}
\end{equation}
Recently this Hamiltonian has been simulated experimentally by atoms in a cavity,
with an exponent tunable in the range $\alpha \lesssim 3$ \cite{Islam2013,exp1,exp2}.

Studying by DMRG calculations the Von Neumann entropy and the energy spectrum of the model described by Eq. \eqref{LRI}, in 
the range of parameters $0<\theta<\frac{\pi}{2}$ (for $\frac{\pi}{2}<\theta<\pi$ the phase diagram is mirrored) 
and $0<\alpha<\infty$, it has been shown 
\cite{koffel2012} that a quantum phase transition separating the 
antiferromagnetic and the paramagnetic phases survives for all finite $\alpha \gtrsim 1$. \\
At variance, below 
this approximate threshold, a new phase arises on the paramagnetic side, 
bounded from above by a transition with non vanishing mass gap and 
preserved spin-flip (along $\hat{x}$ axis) $Z_2$ symmetry (a unique ground-state appears in the DMRG spectrum). 
Correspondingly a logarithmic violation of the area-law for the Von Neumann entropy has been found \cite{koffel2012,paperdouble}. In spite of the limited sizes achieved by DMRG ($L <150$), in \cite{koffel2012} the same violation has been probed also by finite size scaling, showing that it is not originated from finite size effects.

{\color{black}
Notably an exact calculation in the limit $\alpha \to 0$ and $\theta \to 0 , \, \pi$  allows to conclude that the ground state energy of the LR Ising chain is extensive in the LR paramagnetic phase \cite{notext}, so that no Kac rescaling \cite{libro} is required to define rigorously the thermodynamic limit. The same conclusion can be achieved by the study of the Lipkin-Meshkov-Glick (LMG) model \cite{gabbrielli}, coinciding with the LR Ising model when $\alpha \to 0$.
}

The phase diagram for the LR Ising model is depicted in Fig. \ref{phdising}, where the critical semi-line is reported, as well as the violation of the area-law. 
\begin{figure}
\includegraphics[width=0.9\columnwidth]{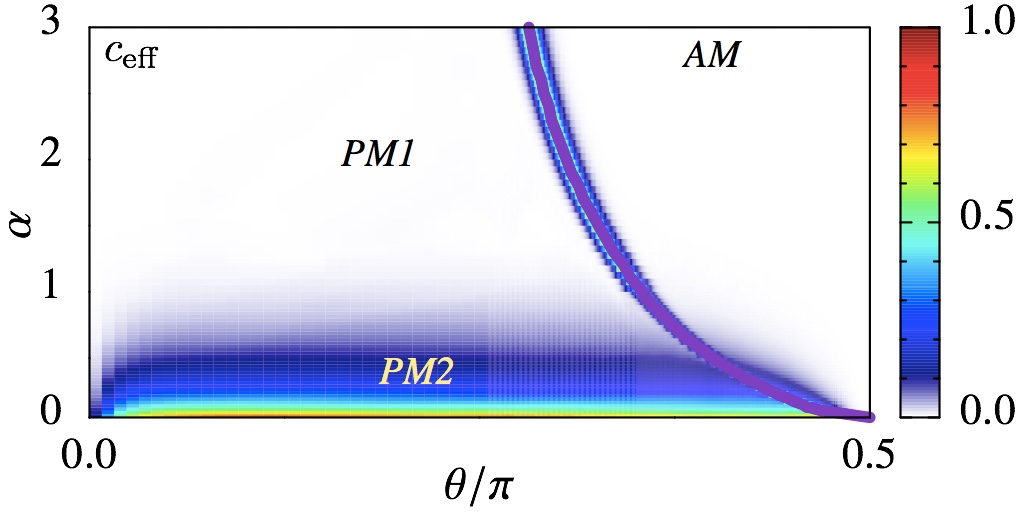}
\caption{Phase diagram of the LR Ising model in Eq. \eqref{LRI}, derived analyzing the area-law deviation for the Von Neumann entropy.
We report in particular the quantity $c_{\mathrm{eff}}$ defined in Eq. \eqref{dev}. The purple semi-line is critical, there the mass gap vanishes. The narrow zone close to the line $\theta =0$ is left white since not investigated, due to a DMRG instability. The antiferromagnetic and the paramagnetic phases at $\alpha >1$ are denoted by the symbols AM and PM1 respectively, while the LR phase on the paramagnetic side at $\alpha <1$ is quoted as PM2. {\color{black} A similar phase diagram has been originally derived in \cite{paperdouble}.}}
\label{phdising}
\end{figure}
Similarly to the LR Kitaev chains, a dynamics of singular states has been recently shown \cite{lepori2016}, responsible for the breakdown of the 
conformal invariance 
along the antiferromagnetic-paramagnetic quantum phase transition at small enough $\alpha$ (from a critical 
$\alpha$ whose value is in the range $1 \leq \alpha \leq 2$ \cite{koffel2012,paperdouble}).

{\color{black}
The LR Ising model, Eq. \eqref{LRI}, can be mapped by means of a Jordan-Wigner transformation
to an interacting LR interacting fermionic chain \cite{paperdouble,noteJW}.
The regime of the latter Hamiltonian at $\alpha \lesssim 1$ corresponding to the paramagnetic one for the LR Ising chain is characterized by the appearance of massive edge states, similar to the ones found for the LR paired Kitaev chain. This facts parallels and confirms the existence of a new phase for the LR Ising model at $\alpha<1$. 
}

\subsection{Entanglement spectrum and inapplicability of {\color{black} MPS-based bulk-boundary} classifications}
\label{isb}
\begin{figure}
\includegraphics[width=0.5\textwidth-10pt]{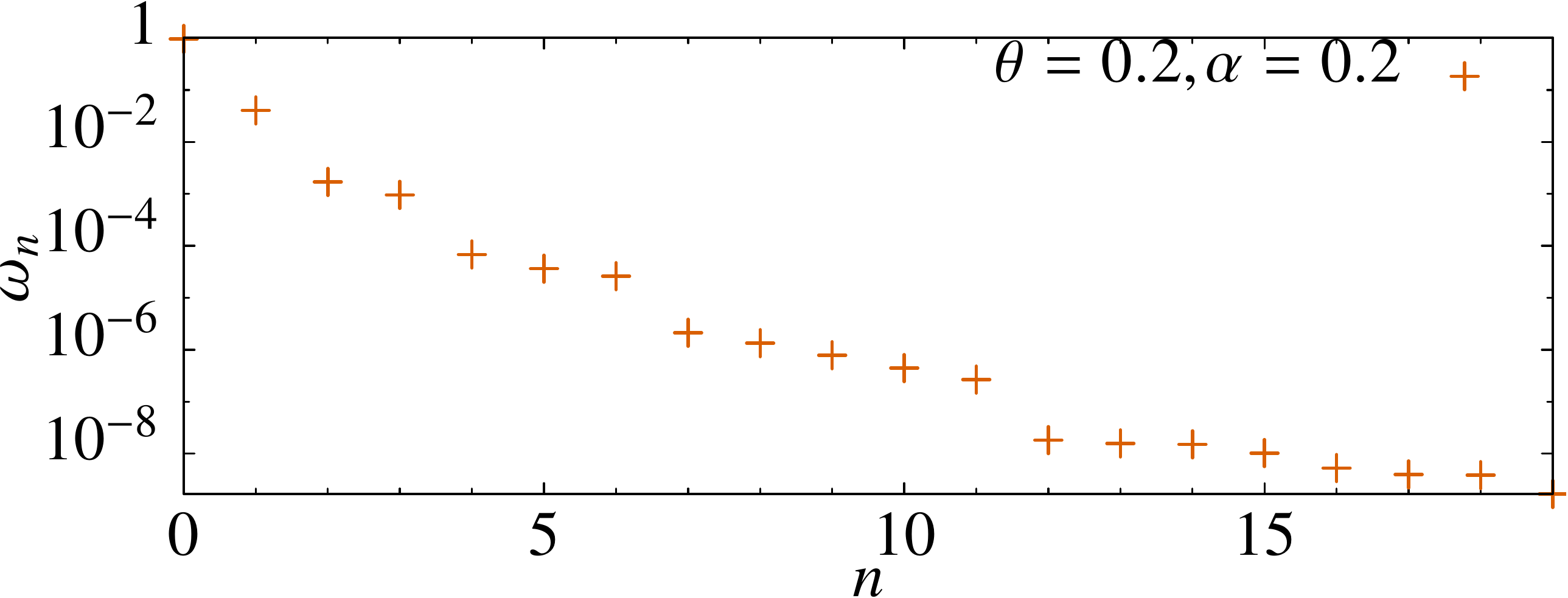}
\includegraphics[width=0.51\textwidth-10pt]{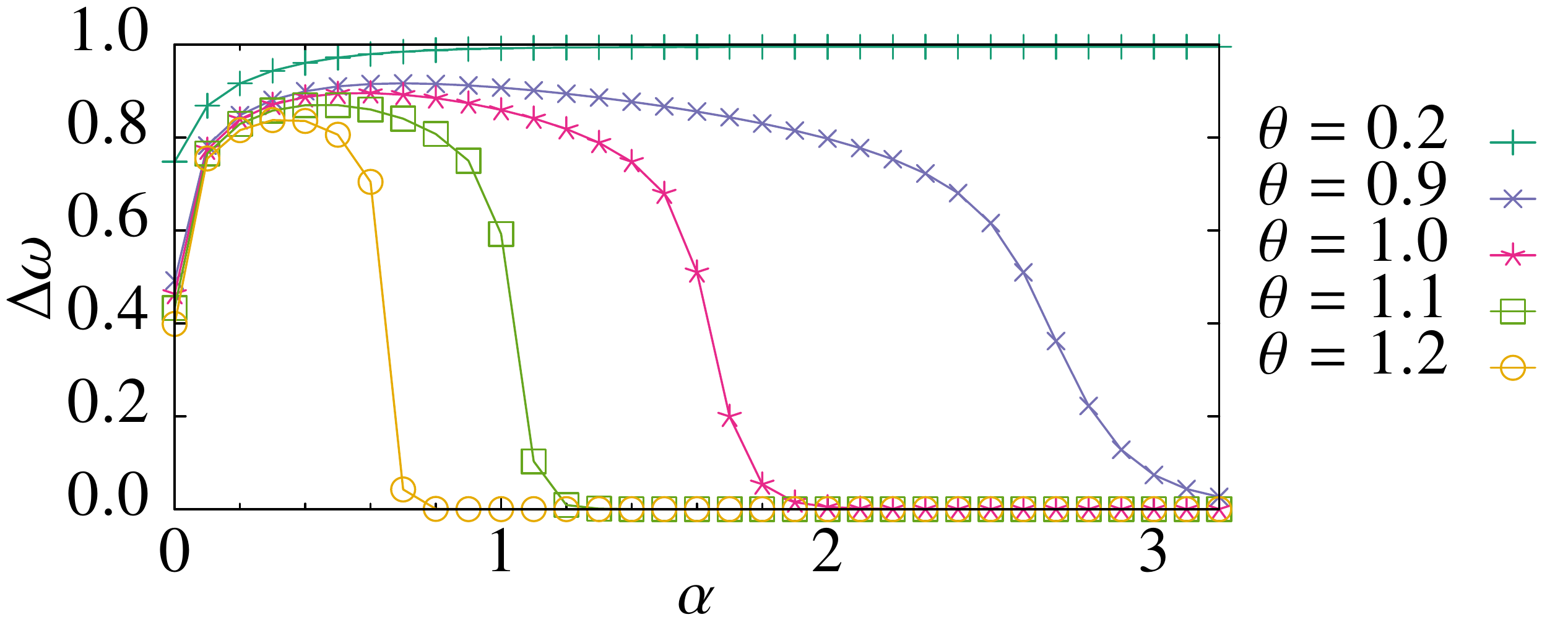}
\caption{Upper panel: distribution of the ES (Schmidt eigenvalues) for the LR Ising chain with $L = 100$, in the point belonging to the LR phase
 at $\theta = \alpha =0.2$. Lower panel: behaviour of the Schmidt gap $\Delta \omega$ as a function of $\alpha$ and for different $\theta$. Notice that $\Delta \omega$ is always nonvanishing along the line $\theta = 0.2$, since the same line never enters in the paramagnetic side where the LR phase occurs.
 Moreover, approaching $\theta= 0$, $\Delta \omega$ increases, as expected.}
\label{ESLRI}
\end{figure}
In this Subsection we study the ES for the LR Ising chain after an half-chain cut and we compare it with the results in the SR limit. We also assume open boundary conditions.

 Exploiting a MPS based boundary operator approach, similar to the one discussed in Section \ref{failES} for the BDI one-dimensional Hamiltonians, 
in \cite{pollmann2010} it was shown that for a purely $Z_2$ symmetric {\color{black} and SR} spin chain only two disconnected phases with preserved $Z_2$ spin-flip symmetry can be found, having the ES contents qualitatively equal to the two phases of the SR Ising Hamiltonian \cite{peschel2011}
 \cite{noteising}) or the open SR Kitaev chain (see Section \ref{failES}).  More in detail, 
the ordered phase displays a distribution for the ES {\color{black} in the form of} Schmidt multiplets with even degeneracy, while the disordered 
phase has no constraint {\color{black} on the ES distribution}, in particular the minimum multiplet degeneracy is equal to 1.\\
{\color{black} We check now if and to what extent the picture described above for the SR quantum Ising chain still remains valid in the presence of 
LR Hamiltonian terms.

We find that the distribution for the ES, typical of the SR disordered phase, occurs also} in the LR phase below $\alpha =1$ on the paramagnetic 
side, as visible in Fig. \ref{ESLRI}. In particular, in the lower panel, 
we display the behaviour of the Schmidt gap (as defined in Section \ref{ES}) $\Delta \omega$ for $L= 100$ and different values of $\alpha$, 
finding a non closure for it if $\alpha<1$, up to finite-size effects. 
We also see that  $\Delta \omega$ at $\alpha \to 0$ increases as $\theta$ gets closer to 0 and the LR phase gets far from its bound with the anti-ferromagnetic phase.

For this reason, similarly to the LR Kitaev chains, we find that the ES distribution found in the LR phase for $\alpha \lesssim 1$ escapes the SR classification \cite{pollmann2010}: indeed again we find, a second (LR) phase with realized $Z_2$ spin flip symmetry and no constraint on the Schmidt multiplets. This behaviour remains up to the LMG limit obtained for $\alpha \to 0$. Since the main steps in the discussion of \cite{pollmann2010} follow the same logic as in \cite{turner2011}, from Section \ref{failES} we can conclude that the reason of this deviation is again the violation (even if only logarithmic) of the area-law for the Von Neumann entropy and the related effective divergence of the correlation length, which 
spoil the MPS-based edge operator approaches.

\section{Conclusions}
\label{conclusions}

In this paper we {\color{black} investigated} the emergence of new types of bulk insulators {\color{black} and superconductors} in the presence of  long-range Hamiltonian terms,
not included in the  classification of the short-range topological insulators {\color{black} and superconductors}, the so-called "ten-fold way classification". 
The reasons of these deviations are analyzed, first studying specific one-dimensional examples, later on focusing on the general structure, in any dimension, of long-range non interacting fermionic Hamiltonians.\\
 The new phases are found originated from 
 a particular type of divergences occurring in the thermodynamic limit due to the long-range couplings: if the latter ones are important enough,   the same divergencies spoil some continuity hypothesis (mainly on the bulk energy spectrum) at the basis of the ten-fold way classification, determining its breakdown.  Related to this fact, a topology can be still defined,  at least for one dimensional systems,  by {\color{black} winding numbers} or Berry phase approaches, provided a proper identification of the long-range contributions to these numbers and consequently of the topologically trivial phases.

From a many body point of view, the central ingredient for the appearance of purely long-range insulating {\color{black} or superconducting} phases 
seems to be the violation of the area-law for the Von Neumann entropy {\color{black} in gapped regimes}. Notably a logarithmic (soft) violation is 
suggested already sufficient for the one-dimensional cases considered in this paper. In these models, the emergence of the long-range regimes deeply 
reflects also on the behaviour of another entanglement indicator, the entanglement spectrum, whose analysis also allowed us to reconsider critically the link between bulk and edge dynamics. 
Moreover, also in the light of the hybrid (exponential plus algebraic) decay behaviour found previously for the static correlation functions, 
the area-law violation induced to re-discuss the concept of correlation length in long-range systems.\\
The stability of the long-range phases against finite-size effects and local disorder is also discussed, showing notably that
current trapped ion techniques should be already able to reach sufficient system sizes to guarantee the observation of the described LR effects.
Moreover we find notably that disorder can even strengthen the effects of the long-range Hamiltonian terms, instead of smearing them, as it could  be naively expected. \\
Concerning the edge properties of the purely long-range phases, the analysis of the entanglement spectrum strongly suggested the partial {\color{black} loss of validity}, at least in one-dimension, of the bulk-edge correspondence, valid instead for short-range topological insulators {\color{black} and superconductors}, due to the strong correlations between bulk and edges dynamics. 
However, the parallelism between the appearance of massive edge states and of nontrivial Berry phases and winding numbers (although defined with particular caveats, such to identity properly the long-range contributions) suggested the emergence of a weakened form of bulk-boundary correspondence, peculiar of long-range quantum systems: {\color{black} there a nontrivial topology still reflects in the presence of states localized on the edges, but these states have a nonzero mass and consequently a dynamics not separable from that of the bulk (in the sense that no modes localized on a single edge can be defined from these bulk states), as it happens instead in the short-range limit}.\\
Finally, the possible extension of some results and ideas for the free long-range insulators {\color{black} and superconductors} has been probed
 on a paradigmatic example of spin model, the long-range Ising chain, which can be mapped to LR interacting fermions. Again important deviations from the structure expected for the short-range spin chains are identified in the entanglement content; consequently the limitations of bulk-boundary (tensor-network based)
 approaches to classify long-range spin models is also discussed.

 Natural developments of the present work are:
\emph{i)} the identification and classification of possible long-range topological phases for quantum systems with arbitrary dimensionality. {\color{black} The same phases are expected since long-range Hamiltonian terms are able to induce (second-type)
divergences in the quasiparticle spectrum no matter the dimensionality of the system.} The analysis of the conditions for the emergence of massive edge states seems to be a promising approach;
 \emph{ii)}  the corresponding investigation of the nature of the edge states in long-range fermionic systems with dimensionality bigger than one, 
 in order to probe the weakened bulk-boundary correspondence (also following the logic in \cite{niu1985,arovas1988}).  
  Interestingly from the experimental and technological points of view, this issue also concerns the possible absence of edge conductivity, present instead for short-range topological insulators {\color{black} and superconductors}. First examples have been recently given in \cite{viyuela2017,lepori2017-2};
 \emph{iii)} the generalization to interacting long-range models, also exploiting entanglement indicators (for instance, in the short-range limit the entanglement spectrum is proved to be effective also in the presence of explicit interactions \cite{turner2011});  
 \emph{iv)}  the study of the effects of stronger deviations from the area-law for the Von Neumann entropy, for instance assuming a (almost) volume-law scaling, as in the set of systems investigated in \cite{gori};  
 \emph{v)} the understanding  of the role of disorder on the singular dynamics in interacting long-range systems, as for the long-range Ising model. There effects
of many-body localization \cite{huse2013} are expected to play a relevant role;
 \emph{vi)} the identification of a general scheme for the experimental detection of the long-range phases, for instance based on direct measurements of the entanglement or of some topological invariants, e.g. by imaging techniques;  \emph{vii)} the study of the stability of long-range 
(topological) phases {\color{black} at  finite temperature.}\\

{\bf Acknowledgements --} 
We thank in a special way A. Trombettoni for the stimulating discussions at the beginning of the work and for the careful reading of the manuscript, A. Turner for the enlightening correspondence,
and D. Vodola for the collaboration at the early stage of this project.
The authors also acknowledge very useful discussions with M. Burrello, 
G. Campagnano, M. Gabbrielli, D. Giuliano, G. Gori, A. Gorshkov, S. Paganelli,  L. Pezz\'e, G. Pupillo, L. Salasnich, and A. Smerzi.\\
Part of this work has been
performed during the  workshop "Conformal Field Theories and Renormalization Group Flows in Dimensions $D>2$"
in the Galileo Galilei Institute for Theoretical Physics, Firenze, 23th May - 8th July 2016.\\

\makeatletter 
\makeatother

\appendix

\section{Long-range paired Kitaev chain}
\label{kitaevapp}

In \cite{nostro,paper1,lepori2016} a quadratic quantum model involving  spinless fermions
on a one-dimensional lattice have been studies extensively. This is 
characterized by a long-range (LR) pairing:
\begin{equation}
\begin{split}
H_{\mathrm{lat}} & = - w \sum_{j=1}^{L} \left(a^\dagger_j a_{j+1} + \mathrm{h.c.}\right)  - \mu \sum_{j=1}^L \left(n_j - \frac{1}{2}\right) 
\\ &+\frac{\Delta}{2} \sum_{j=1}^L \,\sum_{\ell=1}^{L-1} d_\ell^{-\alpha} \left( a_j a_{j+\ell} + a^\dagger_{j+\ell} a^\dagger_{j}\right) \, .
\label{Hamsec}
\end{split}
\end{equation}
For a closed chain, we define in Eq.~\eqref{Hamsec} $d_\ell = \ell$ ($d_\ell = L-\ell$) if $\ell < L/2$ ($\ell > L/2$) and 
we choose anti-periodic boundary conditions \cite{nostro}.\\
The spectrum of excitations is obtained via a Bogoliubov transformation and 
it is given by \big($\omega = \frac{\Delta}{2} \equiv 1$\big):
\begin{equation}
\lambda_{\alpha}(k_n) = \sqrt{\left(\mu - \cos{k_n} \right)^2 + f_{\alpha}^2(k_n + \pi)} \, .
\label{eigenv}
\end{equation}
In Eq.~\eqref{eigenv}, $k_n=- \pi + 2\pi \left(n +  1/2\right)/L$, 
with $0 \leq n< L$ and
$f_{\alpha} (k) \equiv \sum_{l=1}^{L-1} \sin(k l)/d_\ell^\alpha$. For sake of simplicity, in the following the 
subscript $n$ will be neglected.
The functions $f_\alpha(k)$ can be also evaluated in the thermodynamic limit, 
where they become polylogarithmic functions \cite{nist}. \\
The ground-state of Eq.~\eqref{Hamsec} is 
given by $|{\mathrm{GS}} \rangle =\prod_{n=0}^{L/2-1} 
\left(\cos\theta_{k} - i \sin\theta_{k} \, a^\dagger_{k} 
a^\dagger_{-k} \right) |0\rangle$, 
with $\tan(2\theta_{k}) = -f_{\alpha}(k+ \pi)/(\mu -\cos{k} )$; notably it is even under the $Z_2$ parity symmetry of the fermionic number (see below).
\begin{figure}
\includegraphics[width=1.0\columnwidth]{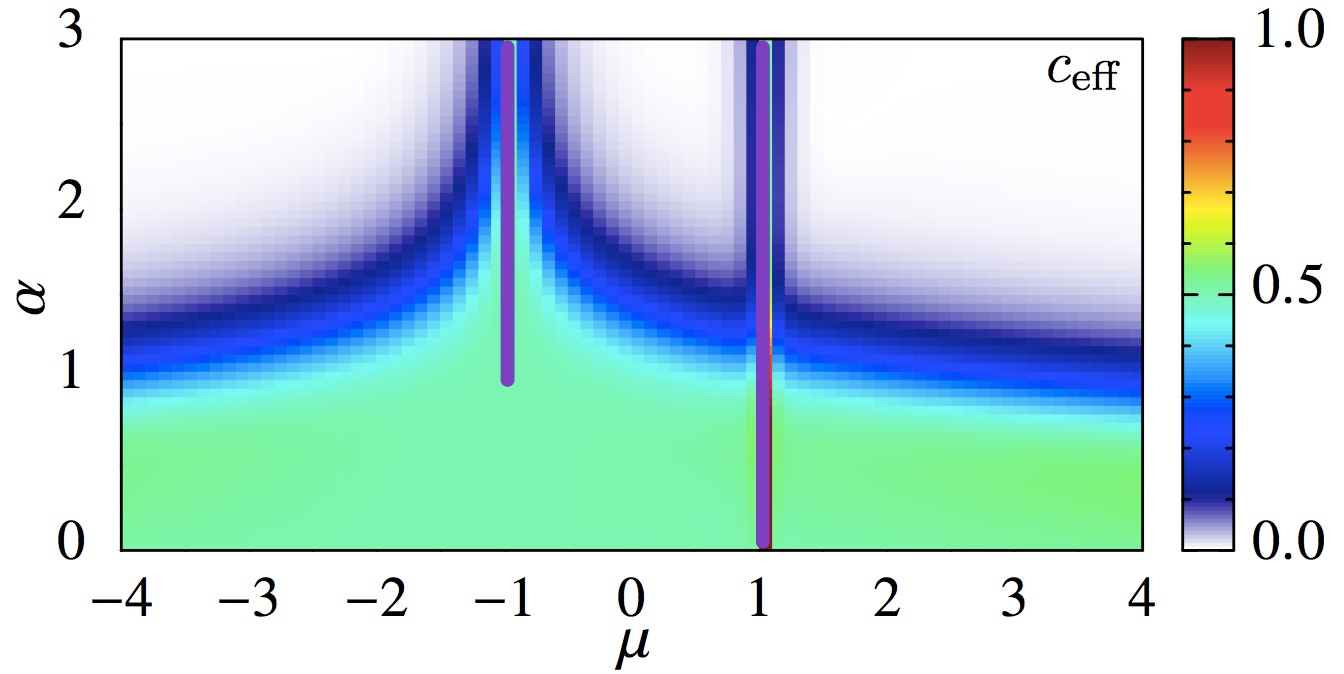}
\caption{Phase diagram of the LR  paired Kitaev chain in Eqs. \eqref{Ham} and \eqref{Hamsec}, derived analyzing the area-law deviation for the Von Neumann entropy.  We report in particular the quantity $c_{\mathrm{eff}}$ defined in Eq. \eqref{dev}. The purple (semi-)lines are critical, there the mass gap vanishes, moreover $c_{\mathrm{eff}} = \frac{1}{2}$ at $\alpha >1$ and $c_{\mathrm{eff}} = 1$ at $\alpha <1$ \cite{paper1}.}
\label{phdkitaev}
\end{figure}

The Hamiltonian in Eq. \eqref{Hamsec} is  invariant both under time reversal symmetry and particle-hole symmetries, realized respectively
by the anti-unitary transformations $\mathcal{U}_T = K \, U_T$ and $\mathcal{U}_c = K \, U_C$, being $K$ the complex-conjugation operator and $U_T^2 = U_C^2 =1$. In this way, it belongs to the BDI symmetry class of the TWC \cite{zirnbauer1996,zirnbauer1997,ludwig2008,kitaev2009,ludwig2009,ludwig2010}. 

The phase diagram of the Hamiltonian in Eq. \eqref{Ham}
 is reported in Fig. \ref{phdkitaev}, plotting the area-law violation for the Von Neumann entropy, quantified as described in the main text. The critical (semi)-lines are also drawn.
 
Further features of the SR limit $\alpha \to \infty$ and the LR regime $\alpha \leq1$
are discussed in more detail in the subsection below.

\subsection*{Known features of the SR limit $\alpha \to \infty$}

In the SR limit $\alpha \to \infty$ the Hamiltonian in Eq. \eqref{Hamsec} reduces to the usual Kitave chain \cite{kitaev}. This model hosts two phases \cite{notetwophases}, characterized respectively by the winding numbers $w=0$ and $w=1$ of the first homotopy class \cite{nakahara,volovik} of the map {\color{black} $k \to \b{n}(k) $, where $\b{n}(k) $ is such that the matrix Hamiltonian in momentum space is written as $H(\b{k}) = |\b{n}_{\b{k}}| \, \hat{n}_{\b{k}} \cdot \vec{\sigma}$ and $\hat{n}_{\b{k}} = \frac{\b{n}_{\b{k}}}{|\b{n}_{\b{k}}|}$ (see the main text).}
In the (disordered) phase with $w = 0$ a unique ground-state $|\mathrm{GS}\rangle$, eigenstate of the $Z_2$ fermionic parity (with even parity), occurs (see e.g. \cite{fendley2012}). This parity is defined in general on the number of fermions $\braket{\hat{F}}= \braket{\sum_{i=1}^L \, a_i^{\dagger} a_i}$ in a certain state.\\
 The second (ordered) phase with $w=1$  is characterized by the presence of two Majorana (massless) edge modes at its ends, exactly due to its nontrivial topology.
  Thanks to the presence of the massless edge modes, two ground-states, $|\mathrm{GS}\rangle$ (defined just above) and $|\mathrm{GS}\rangle_o$, are present in the thermodynamic limit, having different  $Z_2$ fermionic parity.  In particular 
 $|\mathrm{GS}\rangle_o =  \eta_0^{\dagger} \, |\mathrm{GS}\rangle$ has odd parity; the fermionic operator $\eta_0 = \eta_L + i \, \eta_R$ is constructed by the ones related with the two massless (Majorana) edges modes $\eta_{\{R, L\}}$. The states $|\mathrm{GS}\rangle$ and $|\mathrm{GS}\rangle_o$ are degenerate in energy in the thermodynamic limit, exactly because the edge modes are massless.
 {\color{black} However, no spontaneous symmetry breaking, indicated by a local order parameter, occurs (see e.g. \cite{fendley2012,turner2011}).\\}
The two phases also correspond, via Jordan-Wigner transformation, with the ones of SR Ising model, discriminated by (the modulus of) the expectation value of the average longitudinal magnetization $|\langle \sigma_x \rangle |= 	\lim_{l \to \infty} \sqrt{|\langle \sigma_i^{(x)} \sigma_{i+l}^{(x)} \rangle|}$, {\color{black} a local parameter \cite{refpar,notemagn}}. In turn this parameter signals the behavior of the two phases under  the $Z_2$ ($\sigma_x$) spin-flip symmetry, in the two cases realized and spontaneously broken respectively.\\
The spin-flip $Z_2$ symmetry  and the $Z_2$ fermionic parity of the open SR Kitaev chain are related by the following
relation \cite{fendley2012}:
\beq
(-1)^{\hat{F}}  = \prod_{i=1}^L \, \sigma_i^{(x)} \, .
\label{twosym}
\eeq

{\color{black}

\section{Finite size stability of the LR phases at $\alpha <1$}
\label{finite}
In the main text we commented that the deviations from TWC occurring in LR quantum phases of the models in Eqs. \eqref{Ham} and \eqref{Ham2}  are due to the action of the singular states at $k = \pm \pi$, where the divergences in the spectrum $\lambda (k)$ appear for $\alpha <1$. 
A natural general question at this points is how and to what extent the possible LR phases escaping the TWC in the thermodynamic limit can also 
occur in finite-size LR systems, where the divergences encoded by the singular states are smeared. This stability is clear in the analysis of the 
previous Sections, where numerical data for finite chains are reported, however a more formal justification would be desirable.  The same question 
is also relevant for current experiments on LR systems, realized by trapped ions arrays, where very limited sizes (30-40 sites) can be reached 
(see e.g \cite{exp1,exp2}).

The finite-size stability of the LR phases of the Hamiltonian in Eq. \eqref{Ham}  can be understood for instance analyzing the behaviour for different values of $L$ of 
the quantity 
\beq
m_{\alpha} (\mu) = \lim_{R\to\infty}\sqrt{ \mathrm{det} \, G_{R,0} (\alpha , \mu)} \, ,
\label{magn}
\eeq
where
\beq
 \mathrm{det} \, G_{R,0} (\alpha , \mu) = \mathrm{det} \Big[\delta_{R,0} + 2 \braket{GS|a^\dag_R a^\dag_0+a^\dag_R a_0|GS} \Big].
 \label{magn2}
\eeq
This parameter characterizes when 
$\alpha \to \infty$ the paramagnetic-(anti-)ferromagnetic
quantum phase transition of the SR Ising model. Indeed in the same limit
$m_{\alpha}(\mu)$  
coincides \cite{refpar} with the modulus of the average longitudinal magnetization 
\beq
|\langle \sigma_x \rangle| = 	\lim_{l \to \infty} \sqrt{|\langle  \sigma_i^{(x)} \sigma_{i+l}^{(x)} \rangle|}
\label{magnet}
\eeq
 of the SR Ising model: in particular it  has non vanishing values when $|\mu| <1$ only.
The same identification holds at finite $\alpha$,  where $\sigma_i^{(x)}$ refer to the nonlocal spin model obtained from the Hamiltonian in Eq. \eqref{Hamsec} by Jordan-Wigner transformation (see e.g. \cite{muss}).
 \begin{figure}[h!]
\includegraphics[width=0.85\columnwidth]{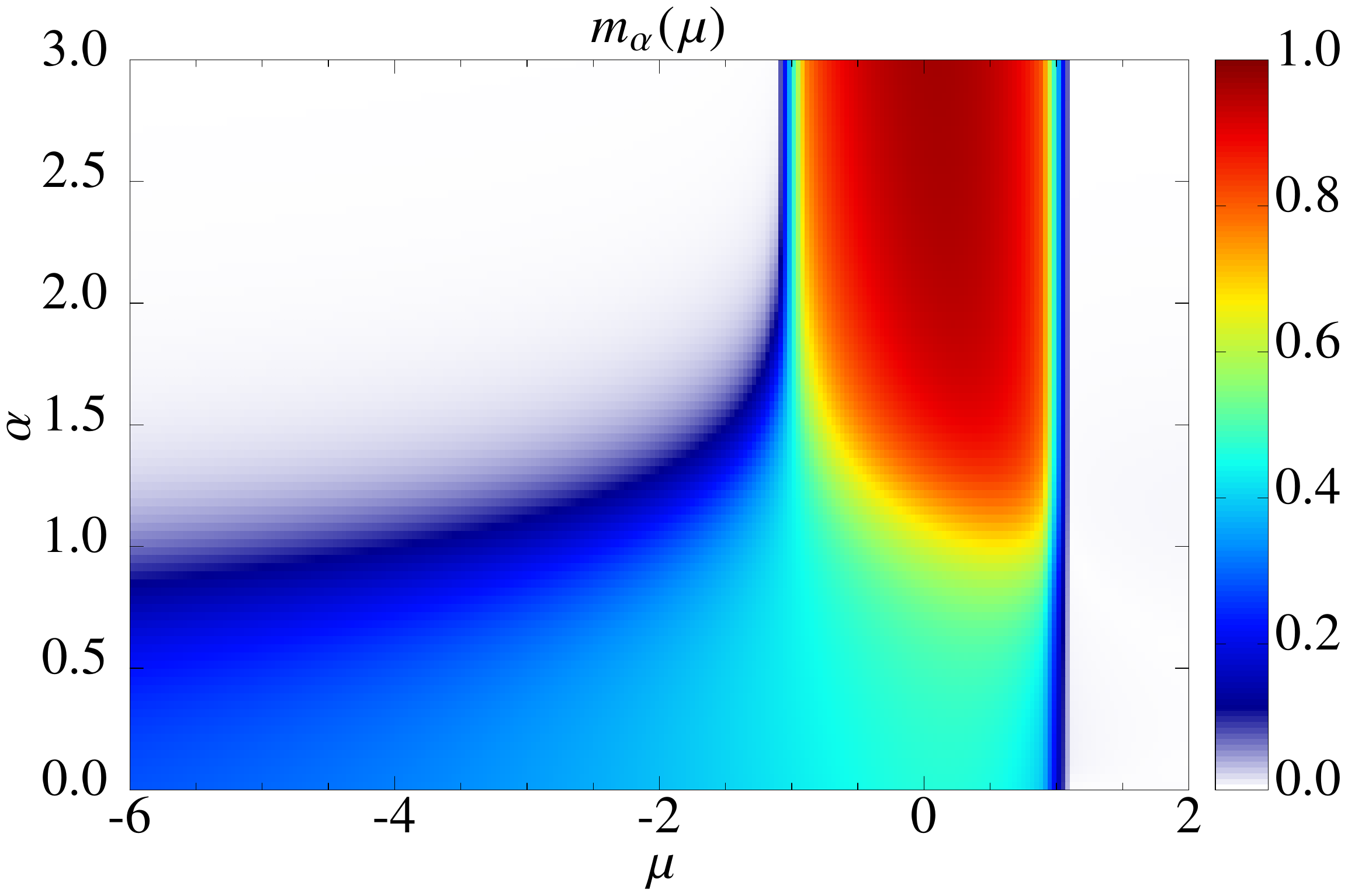}
\caption{Plot of the quantity $m_{\alpha}(\mu)$  in Eq.  (\ref{magn}) for the LR paired Kitaev chain in Eqs. \eqref{Ham} and (\ref{Hamsec}). We assumed $L=800$ and different values of $\mu$ and $\alpha$. Notice that if $\mu >1$, $m_{\alpha}(\mu)$ is found vanishing for every $\alpha$.}
\label{sigma}
\end{figure}

We plot in Fig. \ref{sigma}  the quantity $m_{\alpha}(\mu)$ for $L = 800$, $l = \frac{L}{2} = 400$ and different values of $\mu$ and $\alpha$. There $m_{\alpha} (\mu) $ appears to be  vanishing approximately above $\alpha = 1$ (this threshold being saturated in the limit $L \to \infty$) if $|\mu| >1$, as required for the paramagnetic (disordered) phase proper of the SR limit. On the contrary,
at $\alpha<1$ it assumes a nonzero value  when also $\mu <1$, signaling the outcome of the LR phase with massive edge modes.
It is remarkable that, even at finite $L$,  the zone where $m_{\alpha} (\mu) \neq 0$ coincides within very good approximation  with the zone (at $\mu <1$) where the violation of the area-law for the Von Neumann entropy takes place (see \cite{nostro} for a comparison). 
A similar behaviour for the operator $| \langle \sigma_y \rangle |$, defined in terms of $a_i$ as in \cite{refpar}, can be found at $\mu > 1$ and $\alpha<1$ \cite{pezze2017}.

In Fig. \ref{fig1} we plot instead $m_{\alpha} (\mu)$
at $\mu = -3.2$ and various $L$ and $\alpha$. We find that
 $m_{\alpha} (\mu) \neq 0$ if $\alpha \lesssim 1$ (the extension of the zone where $m_{\alpha} (\mu)$ change value depends on $L$), 
suggesting that the phase with massive edge states survives also in the presence of important finite-size effects, smearing the singular states.\\
Qualitatively the same result is obtained analyzing the mass of the edge states in the regime $|\mu| <1$ and varying $\alpha$ around the line 
$\alpha = 1$ \cite{nostro}.
\begin{figure}[h!]
\includegraphics[width=0.85\columnwidth]{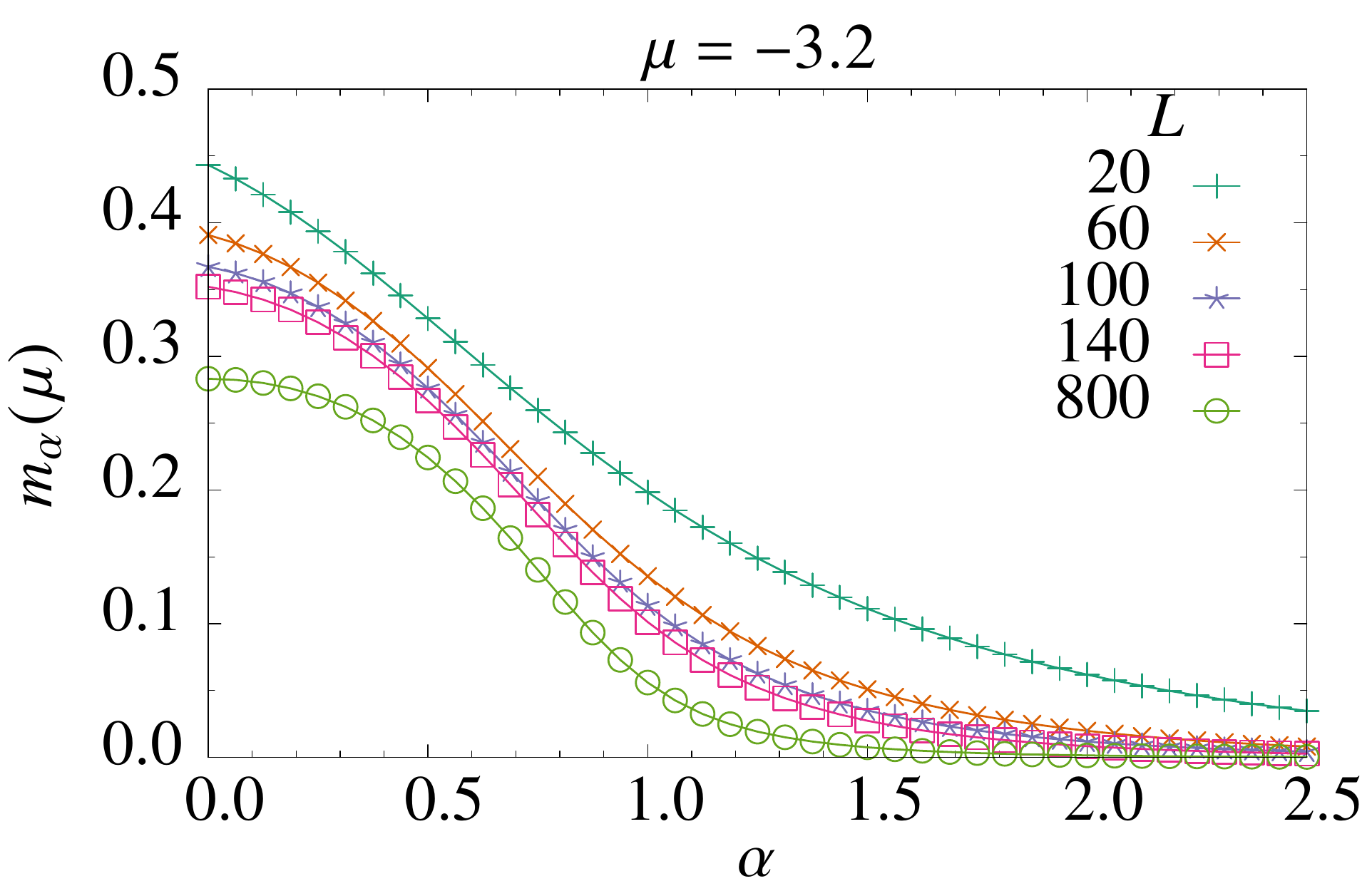}
\caption{Plot of the quantity $m_{\alpha}(\mu)$  in Eq.  \eqref{magn} for $\mu = - 3.2$ and for different values of $L$ and $\alpha$.}
\label{fig1}
\end{figure} 

The behaviour of $m_{\alpha} (\mu)$ can be understood in a better way analyzing how the divergences at $k = \pm \pi$ develop in 
the matrix Hamiltonian in Eq. \eqref{tbHam}, in particular from the contribution due to $ f_{\alpha}(k)$. This can be done following the evolution 
with $L$ and for different $\alpha <1$ of the parameter 
\beq
\mathcal{A} = \frac{| f_{\alpha}(\frac{\pi}{L})- f_{\alpha}(\frac{3 \pi}{L})|}{|\cos (\frac{\pi}{L}) - \cos (\frac{3\pi}{L})|} \, ,
\label{A}
\eeq 
measuring the ratio between the differences  of the functions $f_{\alpha}(k)$ and $\cos(k)$ calculated in the closest point to 
$k = 0$ ($k=\pi$, for the shifted value $f_{\alpha}(k+\pi)$ appearing in Eq.~(\ref{tbHam})), that means $k = \frac{\pi}{L}$, and in the second closest one,  $k = \frac{3 \pi}{L}$. We see in Fig. \ref{ratio} that, at fixed $L$, $\mathcal{A}$  rapidly increases as $\alpha$ decreases from 1 and, even in the most unfavourable case $\alpha \to 1$, we obtain $\mathcal{A} = 10$ if $L \approx 40$. The same threshold value for $\mathcal{A}$ is obtained approximately at $L = 20$ if $\alpha =0.5$.\\  This behaviour means that for every $\alpha <1$ the singularity in $\lambda(\pi)$ develops very rapidly with $L$ increasing, making effective, already at limited sizes, the singular dynamics at the base of the purely LR phases.  \\
\begin{figure}[h!]
\includegraphics[width=0.38\textwidth-10pt]{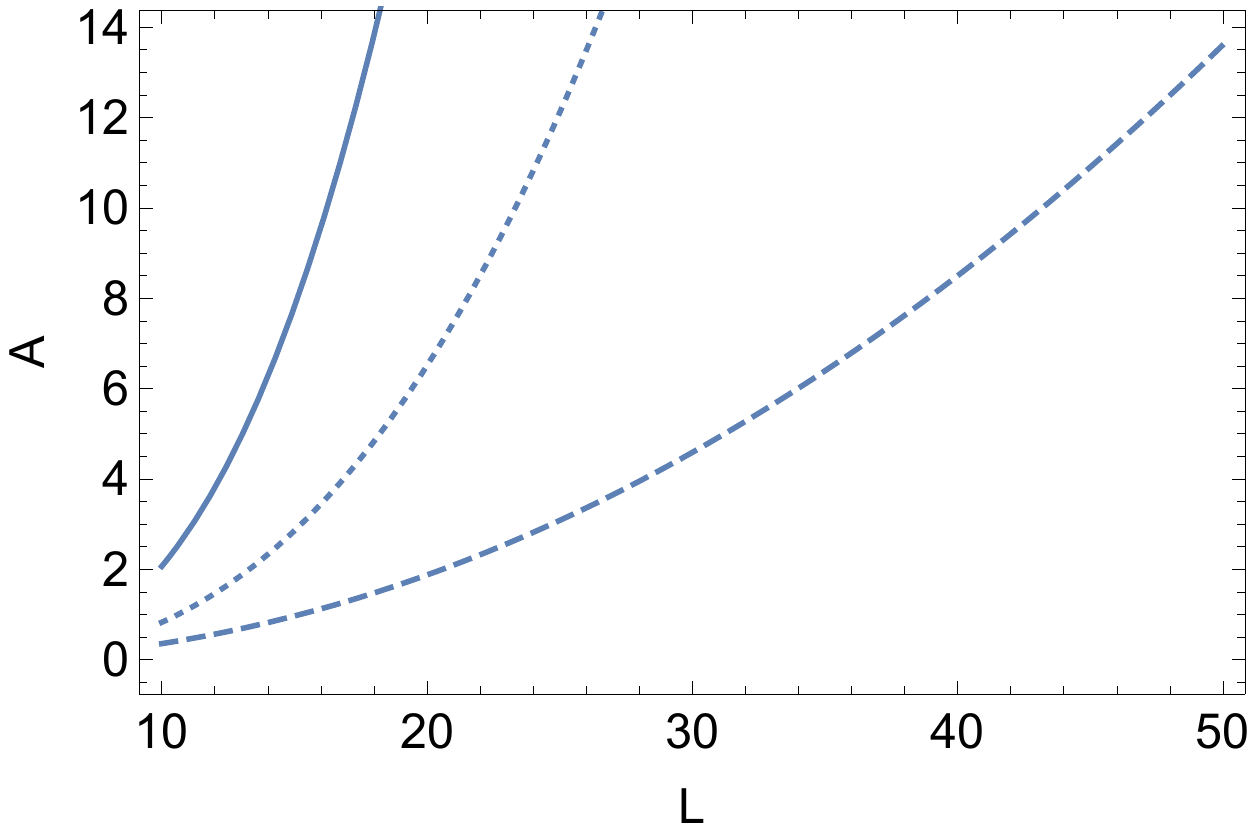}
\caption{Plot of the ratio $\mathcal{A}$ in Eq. \eqref{A} for $L$ varying and $\alpha = 0.99$ (dashed line), $\alpha = 0.5$ (dotted line), and $\alpha = 0.2$ (continuous line). }
\label{ratio}
\end{figure}

The stability of the phases below $\alpha = 1$ can be also clarified on the basis of general considerations on the static correlation functions.
All these quantities can be constructed from the two point correlators 
$g_1 (R) \equiv \langle a^\dag_R a_0 \rangle$ and 
$g_1^{\text{(anom)}}(R) \equiv \langle a^\dag_R a^\dag_0\rangle$
by Wick's theorem.  Their qualitative behaviour, characterized by the typical separating distance $R^*$, has been recalled in Section \ref{correlation}.

We plot in Fig. \ref{corr} (left panel) the behaviour of $g_1 (R)$, for $\mu = -5$, $\alpha = 1.5$ and  $\alpha =3$, and for various system sizes $L$. We see that, decreasing $L $ from $L = 300$, the algebraic tails become shorter and shorter, while the exponential part remains practically stable, as well as the point $R^{*} \approx 20$. Therefore, when $L$ reaches the length $L \approx R^{*}$, the algebraic tail disappears and only the exponential part remains, as in the case of SR systems. A qualitatively same behaviour is found for $g_1^{\text{(anom)}}(R)$. \\
Conversely, decreasing $\alpha$ at fixed $L$, also $R^*$ decreases \cite{nostro,paperdouble,paper1}. In particular, as visible in Fig. \ref{corr}, 
$R^*$ becomes very small in comparison with $L$, so that the decay is purely algebraic. 
The described behaviour holds qualitatively no matter the values of $\mu$ and $\alpha$ and it has been probed also for other LR models (e.g. the LR Ising model \cite{paperdouble}).

From the discussion above, it turns out that the size $R^*$ gives the natural scale for the appearance of the LR physics. 
This means that, when $L\lesssim R^*$, the system, even if described by an Hamiltonian with LR terms, is practically indistinguishable 
from its SR counterpart and speaking about LR physics has no meaning in this condition, where $\mathcal{A} \lesssim 1$. \\
The present analysis justifies \'a posteriori the behaviour observed in Fig. \ref{fig1} for $m_{\alpha} (\mu)$, suggesting the stability of the 
regime with massive edge states for $\alpha <1$, up to very small sizes $L>R^* \to 0$. \\
More in general, we can infer that  possible LR phases escaping the TWC  
remain stable at finite-sizes, in spite of the fact that the
origin of the deviations from TWC, the singular dynamics, is mathematically well defined in the thermodynamic limit only. In this way, the same phases are expected to be probable
in current experiments, where only limited sizes are reachable.
\begin{figure}[h!]
\includegraphics[width=0.46\textwidth-10pt]{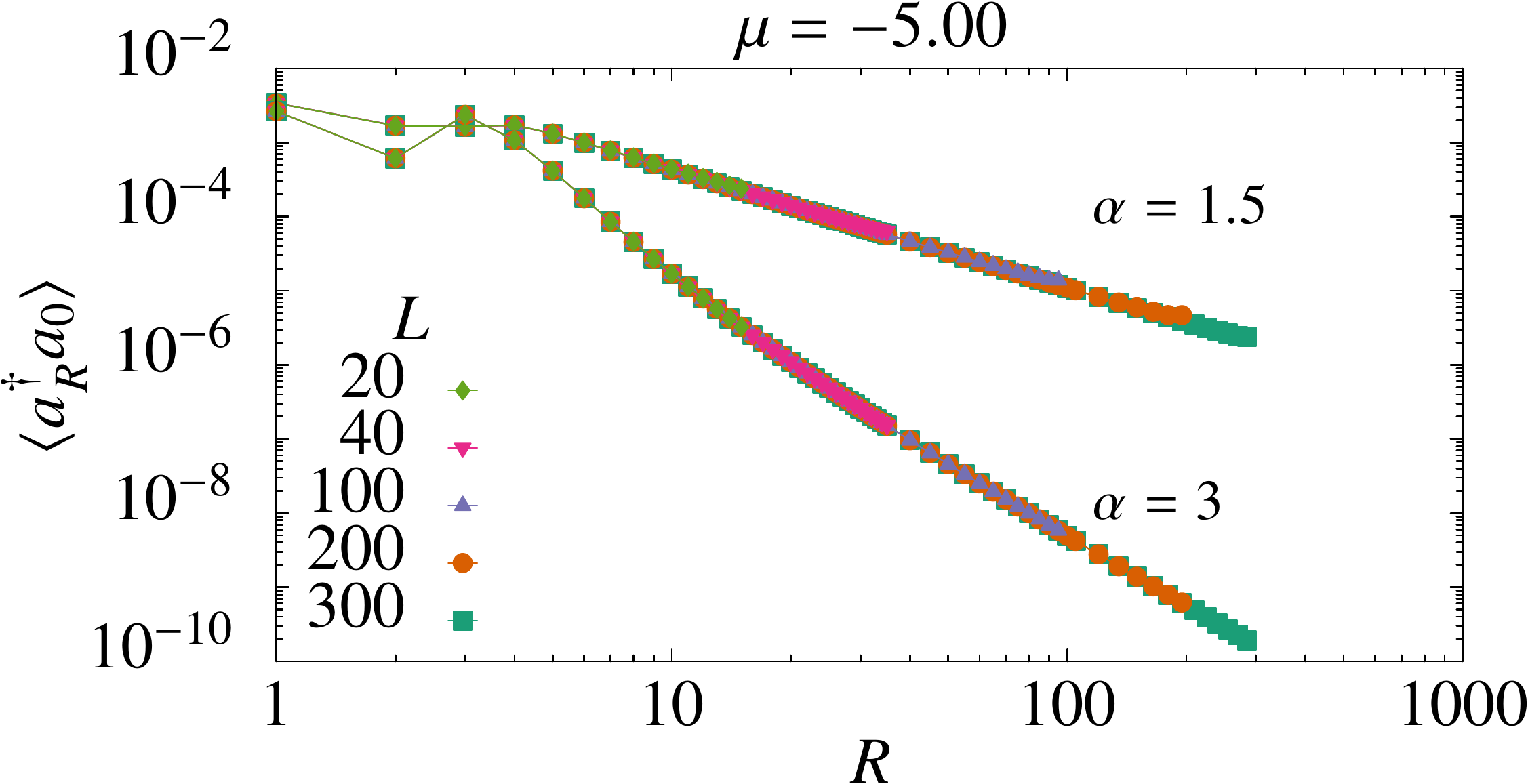}
\caption{Static correlation functions $g_1 (R) \equiv \langle a^\dag_R a_0 \rangle$ in log-log scale for the LR paired Kitaev chain in Eq. (\ref{Ham}), for $\mu = -5$, $\alpha =3$ (lower lines) and $\alpha = 1.5$ (higher lines), and different lengths $L$.}
\label{corr}
\end{figure}
}

\section{Inapplicability of the $\sigma$-model construction}
\label{bulk2}

In this Appendix we show the inapplicability, in the presence of singularities in $H (\b{k})$,
of the nonlinear $\sigma$-model construction leading to the TWC, at least as derived following the
standard approach
\cite{wegner,efetov,belitz,fabrizio,dellanna2006,mirlin2008}.
Here we briefly sketch that derivation without entering too much into 
details and referring to the cited literature for technicalities.

The starting point is the observation that the metal or the insulating nature
of a disordered system is usually described by the behavior of the disorder
averaging of the diffusion propagator \cite{efetov}
\beq
\langle{G_{E+w/2}^R(\b{r},\b{r'})G_{E+w/2}^A(\b{r},\b{r'})} 
\rangle_{\mathrm{disorder}} \, ,
\eeq
where $G_E^{R,A}(\b{r},\b{r'})=\langle\b{r}|(E - H \pm i\eta)|\b{r'}\rangle$ are
the retarded/advanced single particle Green's functions, $\eta \to 0$ is a real
infinitesimal value implementing the usual Feynman prescription (see e.g. \cite{peskin}). 
The total Hamiltonian $H$ contains the free part, here denoted as $H_0$, and  a disorder term
\beq
H_D = \sum_i \epsilon_i  \, a_i^{\dagger} a_i  \, .
\label{dis1}
\eeq
The random variable $\epsilon_i$ is supposed to be  normally distributed:
\beq
P(\epsilon_i)\propto e^{-\epsilon_i^2/4v} \, .
\eeq
Introducing the grassmann variables $\psi$ and $\bar{\psi}$, one can write in the Euclidean space
\[
\frac{1}{E-H\pm i\eta}\propto
\int D\bar{\psi}D\psi\; \psi \bar{\psi} \, e^{-S} \, ,
\]
where $S_0=\int \bar{\psi}(E-H \pm i\eta)\psi$ is disorder dependent.\\
In order to evaluate the effect of disorder on a certain observable  
one should make a stochastic averaging of the quantum expectation values of this observable evaluated at different disorder configurations.\\
 For this purpose, one can resort to the so-called replica method  
\cite{efetov}, which allows to perform disorder averaging 
in terms of quantum expectation values weighted
by a replicated Hamiltonian supplemented by a quartic (interaction) term and taking the zero 
replica limit.
 More specifically, the disorder average of the expectation 
value $\langle {\cal O} \rangle$ 
of a generic operator ${\cal O}$ is given by 
\beq
\overline{\langle {\cal O} \rangle}=\overline{{\Tr(\rho \,  {\cal O})}/{Z}} \, ,
\eeq
 where $Z=\Tr(\rho)$ is the partition function and $\rho$ the density operator which defines the quantum state. \\
Since the random variables are present both in the numerator and in the denominator the stochastic averaging is unpractical. 
However the great advantage of the replica method is that it makes possible to describe the average over disorder of the ratio  
in the form of the ratio of the averages. Indeed, introducing $n$ independent replicas of the system,  we can formally write 
\beq
\overline{\langle {\cal O} \rangle}= \lim_{n\rightarrow 0}{\overline{\Tr \Bigg(\prod_{\alpha=1}^n \rho_\alpha \, {\cal O}_1\Bigg)}}/{\overline{Z^{n}}} \, ,
\eeq 
where ${\cal O}_1$ means that $\cal{O}$ acts only on one replicated system. 
The price to pay is that the effective action acquires a interacting term among the replicas:
\begin{align}
\nonumber \overline{Z^n}=\int D\bar\psi \, D\psi \int d\epsilon_i \, P(\epsilon_i) \, 
\mathrm{exp} \Bigg[-S_0-\sum_{\alpha}\sum_i \epsilon_i \, \bar \psi_{i\alpha}\psi_{i\alpha} \Bigg]\\
=\int D \bar{\psi} D\psi \, \mathrm{exp} \Bigg[-S_0+v\sum_{\alpha,\beta}\sum_{i,j} \bar 
\psi_{i\alpha}\bar\psi_{j\beta} \psi_{j\beta}\psi_{i\alpha} \Bigg] \,,
\end{align} 
where the sum run over the sites ($i$, $j$) and the replica 
($\alpha$, $\beta$) indeces. 
By means of the
Hubbard-Stratonovich transformation, we can decouple this so-obtained quartic term,
introducing an auxiliary matrix field $Q$.
We get therefore an effective action, reading in momentum space:
\begin{eqnarray}
\label{Saction}
S&=&\frac{1}{V}
\frac{1}{{w}} \,{\Tr} \big[Q_{\b{k}}^2 \big] + \\
\nonumber&&+ \sum_{\b{k} , \b{q}} \bar\Psi_{\b{k}} \big\{\left[i \eta \,  s_z +E
- {H}_0(\b{k})\right]\delta_{\b{q},0} + i V^{-1} Q_{-\b{q}} \big\} \Psi_{\b{k+q}}\,.
\end{eqnarray}
The symbol $\Psi_{\b k}$ denotes a multi-spinor in the replica space $\Psi_{\b k}=(\Psi_{\b k 1}, \Psi_{\b k 2}...,\Psi_{\b k n})$, and 
in the particle/hole 
and retarded/advanced ($\pm$) spaces, so that, explicitly, 
$\bar{\Psi}_{\b{k}\alpha}=\left(-{\psi}_{\b{k}\uparrow +}, 
\bar {\psi}_{\b{k}\downarrow +}, 
-{\psi}_{\b{k}\uparrow -}, 
\bar{\psi}_{\b{k}\downarrow -}, 
\right)_\alpha$, 
$\Psi_{\b{k} \alpha}=\left(\bar{\psi}_{\b{k}\uparrow +}, 
{\psi}_{\b{k}\downarrow +}, 
\bar{\psi}_{\b{k}\uparrow -}, 
{\psi}_{\b{k}\downarrow -}\right)_\alpha^t$, 
while $s_z$ is the Pauli matrix in the latter space, {\color{black} $w \propto v^{-1}$ proportional to the scattering time at the Born approximation level}, and $V$ the volume of the space. Let us call {\color{black}  ${\cal G} = \big\{\left[i \eta \,  s_z +E
- {H}_0(\b{k})\right]\delta_{\b{q},0} + i V^{-1} Q_{-\b{q}} \big\}$   } the fermionic 
propagator appearing in Eq.~(\ref{Saction}).\\
Integrating now over the fermionic fields $\Psi_{\b k}$ and $\bar{\Psi}_{\b k}$, one gets an action which depends only on $Q_{\b{k}}$:
\begin{equation}
S[Q]= \frac{1}{V} \,
\frac{1}{{w}} \, {\Tr}\big[
Q_\b{k}^2\big] - \,
\frac{1}{2} {\Tr \, \ln}\, {\cal G}^{-1}\,.
\label{sbos}
\end{equation}
where $\Tr$ is the trace over all the spaces. 
After finding the the saddle point solution $Q_{sp}$, the quantum fluctuations
are such that $Q_{\b{r}}^2=Q_{sp}^2$, and the action can be written as
follows:
\begin{equation}
S[Q]=S[Q_{sp}]-\frac{1}{4}{\Tr \, \ln}(1+G_0 W) \, ,
\end{equation}
where $G_0^{-1}=(E-H_0)^2+Q_{sp}^2$, and
\begin{equation}
W = i \, [Q,H_0] = - \, \b{J} \cdot \b{\nabla} Q \, ,
\end{equation}
where $\b{J} (\b{k})=\nabla_\b{k}H_0(\b{k})$ is a current vertex operator.
In the consequent gradient expansion we also obtain:
\begin{equation}
{\Tr}(G_0 WG_0 W)\simeq {\Tr}(\b{J} \, G_0 \, \b{J} \, G_0) \, {\Tr}
(\b{\nabla} Q \, \b{\nabla} Q) \, ,
\label{expan}
\end{equation}
the desired $\sigma$-model.

The expansion in Eq. \eqref{expan} fails if
$H_0(\b{k})$ diverges {\color{black} (and it is not regolarizable without discontinuities, as for the second-type singularities)}. 
In this condition the charge $\int \mathrm{d} \b{k} \, \b{J}(\b{k})$ also diverges. 
For this reason the $\sigma$-model characterizing the coset $F$ cannot be constructed. 
The present discussion leaves open the possibility of the inapplicability of the $\sigma$-model construction also when only $\b{J} (\b{k})$ 
diverges.

\section{Structure of the edge states at $\alpha <1$}
\label{appedge}

In Section \ref{natedge} we mentioned for the LR Kitav chains
in the Eqs \eqref{Ham} and \eqref{Ham2} the impossibility to identify, in the LR regimes at $\alpha<1$, low-energy states localized separately on the left-hand and the right-hand edges. This impossibility, directly encoded on the ES structure analyzed in Sections \ref{ES} and \ref{failES}, has been claimed {\color{black} in the same Section  \ref{natedge} to be in a one-to-one correspondence with the nonvanishing of the masses for the edge states.}

In order to substantiate our thesis, it is useful to
start  from the construction of the Bogoliubov states
 for quadratic fermionic Hamiltonians. 
 As it happens in the ordered phase for the open SR Kitaev chain (and for the Hamiltonians in Eqs. \eqref{Ham} and \eqref{Ham2}), the fermionic state $\ket{m}$, {\color{black} whose wavefunction is  localized symmetrically at both the edges of the chain,} can be written as \cite{refpar}
\beq
\ket{m} = \eta_{m}^{\dagger} \, \ket{GS} = \sum_{i=1}^L \, \big(g_{m i} \, a_i + h_{m i} \, a_i^{\dagger} \big) \, \ket{GS} \, ,
\label{combmajo}
\eeq
a similar ansatz holding for the other (bulk) eigenstates of the Hamiltonians. 
Notice that, compared to $\ket{GS}$, $\ket{m}$ differs in the fermionic number/parity by a unit; {\color{black} this fact is encoded in the different sign on the two states of the topological pfaffian invariant  discussed in \cite{tewari2012}.\\
As suggested by the linearity of the diagonalizing ansatz for the free Hamiltonians in Eqs. \eqref{Ham} and \eqref{Ham2} and following what done in the SR limit (where $m = 0$), one could attempt to decompose the state $\ket{m}$, involving symmetrically both the edges, 
defining two (right and left) edge operators $f_{R/L}$ as follows (see e. g. \cite{stern2008,paperdouble}):
\beq
\eta_m = \frac{1}{\sqrt{2}} \, \big(f_R + e^{i \phi} f_L \big)
\label{combmaio2}
\eeq  
(the pre-factor $\frac{1}{\sqrt{2}}$ testifying the same weight for the two edges and  $\phi$ being a phase constant to be fixed), depending linearly on $a_i$ and $a_i^{\dagger}$. If $m = 0$, the so constructed operators $f_{R/L}$ ($\phi =\frac{\pi}{2})$, fulfilling the Majorana  condition $f_{R/L}^{\dagger} = e^{i \, \theta_{R/L}} \, f_{R/L}$, are related with two wavefunctions localized separately on each edge.\\
For the Hamiltonians in Eqs. \eqref{Ham} and \eqref{Ham2}, the situation is very different  if $\alpha <1$. Indeed, since  $m \neq 0 $,  the operators $f_{R/L}$ do not fulfill any longer the Majorana  condition, as argued in \cite{paperdouble};
for this reason the canonical anti-commutation rules for $\eta_m$, $\{\eta_m, \eta_m^{\dagger}\} =1$ imply:
\beq
\{f_R, f_R^{\dagger} \} = \{f_L, f_L^{\dagger}\} = 1 \, .
\label{frac}
\eeq
In this way, $f_{R/L}$ are usual fermionic operators, able to induce states \big(as $\ket{R/L}  = f_{R/L}^{\dagger} \ket{GS}$\big) of the Hilbert space for the  considered Hamiltonians. The same possibility does not hold instead if $m=0$, since $\{f_{R/L}, f_{R/L}^{\dagger} \} = \{f_{R/L}, f_{R/L} \} = 0$ and physical states can be constructed only by combinations of them, as in Eq. \eqref{combmaio2} (see for instance \cite{stern2008}).
However, states as $\ket{R/L}$ do not belong to the Hilbert space of the Hamiltonians in Eqs. \eqref{Ham} and \eqref{Ham2}, suggesting that the construction in Eq. \eqref{combmaio2}, although formally possible, is not correct in the absence of the Majorana condition for $f_{R/L}$, then if $m \neq 0$. On the contrary, only the state $\ket{m}$,
involving both of the edges, makes sense in this condition.\\
The other possibility $\ket{m} = f_R^{\dagger} \, f_{L}^{\dagger} \ket{GS} $ is  ruled out by the linearity of the diagonalization problem for the considered quadratic Hamiltonians,  as well as by the fact that the canonical anti-commutation rules $\{a_i , a_j^{\dagger}\} = \delta_{ij}$   allow cancellations of $a_i$ and $a_i^{\dagger}$ only pairwise, then a linear ansatz as in Eq. \eqref{combmajo} cannot be obtained from the  quadratic ansatz for $\ket{m}$ just above. Finally, nonlocal ansatzs are discarded since the beginning, in such a way as not to change the locality property of the excitations in the bulk/edge spectrum (as done by the Jordan-Wigner transformation for the Majorana modes in the SR limit, see e. g. \cite{greiter2014}).


\begin{thebibliography}{XX}
\bibitem{notext}
The extensivity of the ground-state energy of the Hamiltonian in Eq.~\eqref{LRI}  can be shown calculating it in the limit  $\theta \to  \pi$ (even if in this limit a first order transition occurs \cite{vidal2004}). 
A variational calculation shows that in this case 
the ground state can be generally written as a (superposition of classical) state(s) with $\frac{L}{2}$  spins for which the eigenvalues of $\sigma_i^{(x)}$ are equal to $\frac{1}{2}$. The same number of spins, $\frac{L}{2}$, is counted for $\sigma_i^{(x)} = - \frac{1}{2}$. The total energy for 
these states (whose number is $\Big( \begin{matrix} L\\
L/2 \end{matrix} \Big)$, assuming $L$ even) is  $E_{GS} = - \frac{1}{2} \,L \sin \theta$. 
We have therefore that in this limit the ground-state energy is extensive. The same result should be valid for any $\theta$ and $\alpha\geq 0$.
\end{thebibliography}
\end{document}